\documentstyle[emulateapj] {article}

\slugcomment{Submitted to Astronomical Journal.}

\lefthead{Scoville et al.}
\righthead{NICMOS Imaging}

\newcommand{\msun}{\mbox{$M_{\odot}$}}

\newcommand{\lsun}{\mbox{$L_{\odot}$}}
\newcommand{\kms}{\mbox{km s$^{-1}$}}

\def\deg      {{\ifmmode^\circ\else$^\circ$\fi} } 

\def\h2     {H$_2$}

\begin{document}

\title{NICMOS Imaging of Infrared-Luminous Galaxies}

\author{N. Z. Scoville}
\affil{California Institute of Technology, Pasadena, CA 91125}

\author{A. S. Evans}
\affil{California Institute of Technology, Pasadena, CA 91125
\and Dept. of Physics \& Astronomy, SUNY, Stony Brook, NY 11794-3800}

\author{R. Thompson,~~M. Rieke, ~~D. C. Hines, ~~F. J. Low}
\affil{Steward Observatory, University of Arizona, Tucson, AZ 85721}

\author{N. Dinshaw}
\affil{UCO/Lick Observatory, Santa Cruz, CA 95064}

\author{J. A. Surace and ~~L. Armus}
\affil{California Institute of Technology, Pasadena, CA 91125}


\begin{abstract}
We present near-infrared images obtained with the HST NICMOS camera
for a sample of 9 luminous (LIGs: $L_{\rm IR} (8-1000 \mu{\rm m}) \ge
10^{11}$ L$_\odot$) and 15 ultra-luminous (ULIGS: $L_{\rm IR} \ge 10^{12}$
L$_\odot$) infrared galaxies. The sample includes representative systems
classified as warm ($f_{25 \mu{\rm m}}/f_{60 \mu{\rm m}} > 0.2$) and cold
($f_{25 \mu{\rm m}}/f_{60 \mu{\rm m}}\le 0.2$) based on the mid-infrared
colors and systems with nuclear emission lines classified as HII (i.e.
starburst), QSO, Seyfert and LINER. The morphologies of the sample
galaxies are diverse and provide further support for the idea that they
are created by the collision or interactions of spiral galaxies. Although
no new nuclei are seen in the NICMOS images, the NICMOS images do reveal
new spiral structures, bridges, and circumnuclear star clusters. The
colors and the luminosities of the observed clusters are consistent
with them being young ($10^{7-8}$ yrs), formed as a result of galactic
interactions and having masses much greater than those of Galactic
globular. In NGC 6090 and VV114, they are preferentially situated along
the area of overlap of the two galactic disks.

With the exception of IR 17208-0018, all of the ULIGs have at least
one compact ($2.2\mu$m FWHM $\le 200$ pc) nucleus. Analysis of the
near-infrared colors (i.e., $m_{1.1-1.6}$ vs. $m_{1.6-2.2}$) derived
from 1.1$\arcsec$ diameter apertures suggests that the warm galaxies have
near-infrared colors consistent with QSO+hot dust emission and the cold
galaxies, as a group, have near-infrared colors consistent with reddened
starlight. In addition, the cold ULIG UGC 5101 (and possibly three others)
have near-infrared colors suggesting an additional AGN-like near-infrared
component in the nucleus.  In a 2 kpc-diameter aperture measurement,
the global colors of all of the cold galaxies except UGC 5101 are
consistent with starlight with a few magnitudes of visual extinction.
The general dichotomy of the near-infrared properties of the warm and the
cold galaxies are further supported by the light distributions - seven of
the eight warm galaxies have unresolved nuclear emission that contributes
significantly (i.e., $\ge 30-40$\%) to the total near-infrared luminosity.
The smooth, more extended light observed in all of the galaxies is most
likely comprised of giant and supergiant stars, but evidence at longer
wavelengths suggests that these stars contribute little to the high
8--1000$\mu$m luminosity of these galaxies.  Finally, light profiles
of nine of the 24 systems were fit well by an r$^{1/4}$ law (and not
so well by an exponential disk profile).  Whether these star systems
eventually become massive central bulges or giant elliptical galaxies
will depend on how efficiently the present ISM is converted into stars.

\end{abstract}


\section{Introduction}

Luminous infrared galaxies, which emit a substantial amount of their
bolometric luminosities in the wavelength range 8--1000 $\mu$m (e.g.
\cite{rie72}; \cite{soi87}; see Sanders \& Mirabel 1996 for a review),
are observed to be in a phase of dynamically triggered evolution.
Ground-based optical to near-infrared imaging of the most powerful
galaxies detected in the IRAS survey (\cite{soi87}) have revealed that
virtually all show evidence of a strong interaction (eg. extended tidal
tails) or double nuclei (cf.  \cite{jos85}; \cite{arm87}; \cite{san88a};
\cite {cle96}; \cite{mur96}).  Extensive optical and near infrared
spectroscopic surveys have shown that these galaxies can be classified
as luminous starbursts and active galactic nuclei, AGN (\cite{san88a};
\cite{arm89}; \cite{kim95}, 1998; \cite{vei99a}; \cite{vei99b}),
both of which are presumably fueled by the abundance of molecular gas
(\cite{san91}; \cite{sol97}) that collect via gravitational torques
and dissipate into the merging nuclei (\cite{mih96}; \cite{sco97};
\cite{bry96}).  Most of the direct ultraviolet and optical light emitted
by these starbursts and AGN is obscured by the interstellar dust of the
merging galaxy, but is re-emitted at far-infrared wavelengths.

Evidence of morphological disturbances (e.g. Mackenty \& Stockton
1984) and large amounts of molecular gas (e.g. Sanders, Scoville,
\& Soifer 1988) and dust in a few optically-selected QSOs, combined
with the similarities in the space densities of QSOs and ULIGs
(\cite{soi87}), initially led to the hypothesis that ULIGs might
evolve into optically-selected QSOs, once most of their gas and
dust are consumed/blown out by star formation and AGN activity
(\cite{san88a,san88b}).  During this process an accompanying evolution
in the far-infrared color, from cold, starburst-like colors to warm,
Seyfert-like colors, takes place (\cite{san88b}).  The scenario is
supported by optical and mid-infrared spectroscopy, which shows an
increasing fraction of AGN-like emission-line spectra in ULIGs as
a function of increasing luminosity (\cite{vei99a}; \cite{vei99b};
\cite{eva98}; \cite{lut98}).

While the nature of the luminous infrared galaxies are believed
to be understood in general terms, the details have been under
investigation for over a decade. However, the wealth of ground-based
spectroscopic and low resolution imaging surveys have provided little
information on the circumnuclear region at scales comparable to the
dimensions of the starbursts and/or AGN believed to be responsible for the
bulk of their bolometric luminosities.  Optical imaging of these galaxies
has been obtained with Hubble Space Telescope (HST) with resolutions
$<0.1\arcsec$; however, at these wavelengths the circumnuclear regions are
still obscured by the high dust column densities. The recent installation
of the Near-Infrared Camera and Multi-Object Spectrometer (NICMOS) on HST
has now made it possible to observe more directly the embedded nuclear
regions by providing the high resolution possible with HST
(0.1--0.2$\arcsec$, which corresponds to 30-200 pc over the redshift range
0.01--0.1) at wavelengths where the effects of dust extinction are reduced
by an order of magnitude (in exponential optical depth) compared to visual
wavelengths.  We present here an investigation of the near-infrared
morphologies of 24 luminous infrared galaxies using the new capabilities
of HST. The analyses and conclusions drawn from this study benefit from
previous ground-based imaging and spectroscopy of these galaxies.

This paper is divided into eight Sections. Section 1 concludes with a
discussion of the general objectives of the survey and the sample
selection.  In \S 2, the observations, data reduction, and flux
calibration are summarized. Included here is a description of the Point
Spread Function (PSF) subtraction performed on the brighter, unresolved
nuclei and the extinction corrections applied to the 2.2 $\mu$m images.
The morphologies of the individual galaxies are summarized in \S 3 and \S 4,
followed by a discussion of the aperture colors and their intepretation in
\S 5. Section 6 contains a discussion of the radial profile fits to the
images and how well they are fit by elliptical galaxy and exponential disk
profiles. In \S 7, the degree of central compactness is discussed, and in
\S 8 the nature of the extended near-infrared light is considered. Section
9 is a summary of the paper.  Throughout this paper, we adopt H$_0$=75
\kms Mpc$^{-1}$ and q$_0=0$; at the typical redshift (z=0.05) of the
observed galaxies, 0.1--0.2\arcsec\ resolution corresponds to 100--200
pc.

\subsection {General Objectives}

The NICMOS images are particularly well-suited to defining the small-scale
structure of the dust embedded galactic nuclei in infrared luminous
galaxies. The 1.1, 1.6, and 2.2 $\mu$m measurements can be used to
determine the distribution of dust extinction (assuming intrinsic colors
of the embedded stellar populations) and the observed light distribution
corrected for this extinction to better estimate the `true' light
distribution.  The issues we address here include :

1) the location and size of the galactic nucleus (or nuclei) -- an 
   unresolved resolved nucleus (i.e. $\leq 0.1\arcsec$) might
   indicate an AGN or compact, super-luminous starburst -- and
   whether the fraction of the total 2.2 $\mu$m light emitted from compact
nuclear  
   sources is correlated with the total bolometric luminosity or the 
optical
   emission line spectral classification and far-infrared color.

2) the morphology of the dust extinction in the galactic nuclei
   as an indication of central dust accretion disks or rings

3) the existence of inner spirals or bars both of which could transport 
   material to the nucleus and stimulate starbursts in the inner disk

4) the properties and distribution of luminous (presumably young) star
clusters
   formed in the galactic disks and nuclear regions

5) the stellar surface brightness profiles and their relationship to the
stage of dynamical evolution of the galaxies

\subsection {NICMOS Luminous Infrared Galaxy Sample}

 The sample of galaxies range in infrared luminosities from 10$^{11}$ to 4
x 10$^{12}$ \lsun~at $\lambda$=8-1000 $\mu$m.  The sample is not complete
in a flux or distance limited sense, but is instead intended to cover a
broad range of luminosity and intrinsic properties (and presumably
evolutionary stages). The majority of the galaxies were taken from the
IRAS Bright Galaxy Survey which includes all extragalactic objects (324
galaxies) brighter than 5.4 Jy at 60 $\mu$m at $|b|\ge 30\deg$ and
$\delta\ge -30\deg$ (\cite{soi89}). Since one of the primary objectives of
this study was to look for morphological similarities between the
ultraluminous infrared galaxies, AGNs and nuclear starbursts, we have
included for reference two optically-selected QSOs which are also warm
ultraluminous infrared galaxies (3C48 and Mrk 1014) and the relatively
nearby starburst galaxy NGC 6090. We then selected a large fraction of the
most luminous systems, the ultraluminous infrared galaxies (ULIGs) which
have L$ \geq 10^{12}$ \lsun~at $\lambda$=8-1000 $\mu$m.  Nine of the ten
galaxies in the original list of ULIGs ({\cite{san88a}}) are included
here; the remaining object (Mrk 231) was imaged in a companion study by
Hines et al. (1999). We also include here 5 of the 12 objects contained in
the warm ultraluminous sample (\cite{san88b}); these were not in
the BGS due to their lower 60 $\mu$m flux. As stated in the Introduction,
the warm ULIGs have
been suggested as a transition class between the cold ULIGs and
optical/UV excess QSOs.

The galaxies in this sample are listed in Table 1 along with their optical
nuclear emission line and mid-infrared warm (W) versus cold (C)
classifications (with the dividing line between the two classes at a
25-to-60 $\mu$m flux density ratio of 0.20: \cite{deg85};  \cite{deg87};
\cite{low88}; \cite{san88b}), their $\lambda$=8-1000 $\mu$m luminosities,
molecular hydrogen masses (based on mm-CO emission; \cite{san91}) and a
description of the morphology seen in the NICMOS images.
Fig.~\ref{lir2560} shows the sample galaxies plotted as a function of
$L_{ir}$ at $\lambda$=8-1000 $\mu$m and their 25-to-60 $\mu$m flux density
ratios. The classification of the optical emission-line ratios are
indicated by the different symbols.

\section{Observations and Data Reduction}

HST observations of each galaxy were obtained in a single orbit (except
VV114) on the dates listed in Table 2 using camera 2 of NICMOS.  The
camera has a 256$\times$256 HgCdTe array with pixel scales of
0.0762$\arcsec$ and 0.0755$\arcsec$ per pixel in $x$ and $y$,
respectively, providing a $\sim19.5\arcsec\times19.3\arcsec$ field of view
(Thompson et al. 1998).  Images were obtained using the wide-band
 F110W (1.10 $\mu$m, $\Delta\lambda_{\rm FWHM} \sim 0.6$ $\mu$m) and F160W
(1.60 $\mu$m, $\Delta\lambda_{\rm FWHM} \sim 0.4$ $\mu$m) filters, and the
medium-band  F222M (2.22 $\mu$m, $\Delta\lambda_{\rm FWHM} \sim 0.14$
$\mu$m) filter. During five of the orbits, suitably placed HST guide stars
were also imaged for point-spread function (PSF) determinations. The FWHM
are 0.11$\arcsec$, 0.16$\arcsec$, and 0.22$\arcsec$ at 1.1, 1.6, and 2.22
$\mu$m, respectively.  Observations of both the galaxies and the PSF stars
were done using a four or five-point spiral dither in each filter setting;
the step size in most cases was 25.5 pixels (1.91$\arcsec$). The dithered
observing mode yields better sampling of the PSF (due to the half-pixel
steps) and the ability to identify bad-pixels, pixels temporarily affected
by cosmic rays, and processing artifacts. At each dither position,
non-destructive reads (MULTIACCUM) were taken.  The total
integration times for each filter are listed in Table 2. Dark images were
obtained at the end of each orbit.

The initial data reduction and calibration was done with IRAF.  The dark
was first created, then the NICMOS data were dark subtracted, flatfielded
and corrected for cosmic rays using the IRAF pipeline reduction routine
CALNICA (Bushouse 1997).  The calibrated images contained pixels with
reduced quantum efficiency due to contaminants on the array surface, thus
a mask was created to minimized their effect. Two additional areas
required masking; the coronographic hole was masked on all images, as well
as column 128, which is noteably sensitive to minute discrepancies in dark
subtraction (i.e., the dark current rises sharply toward the center column 
of the array).  The dithered images were then shifted and averaged using the
DRIZZLE routine in IRAF (e.g. Hook \& Fruchter 1997). The plate scales of
the final ``drizzled'' images are 0.0381$\arcsec$ and 0.0378$\arcsec$ per
pixel in $x$ and $y$.

The reduced images are shown as 3-color images in Fig.  ~\ref{3color}.
Because of the very large dynamic range and signal-to-noise ratio in the
surface brightness of the galaxies, a variable resolution smoothing
routine (see Appendix A) was applied to the image data for the contour
maps (Fig.  ~\ref{4contour} and ~\ref{2contour}). This routine smooths the
image with a variable width boxcar filter with the resolution depending on
the local signal-to-noise ratio across the image.

 For each galaxy, the three broad-band images were obtained during a
 single guide star acquisition within one orbit. Since the stability of
the guide star tracking for all of these NICMOS observations is much
better than a single pixel, the relative registration of the three images
was assured. On the other hand, in no instance did we attempt absolute
registration of the images using multiple guide stars. We therefore adopt
coordinates in each galaxy measured as offsets from the 2.2 $\mu$m peak,
using the same pixel origin for the 1.1 and 1.6 $\mu$m images. All of the
images are rotated with north up and east to the left using the
data-header orientation angle.

\subsection {Flux Calibration}

Flux calibration of the images were done using scaling factors of
2.031$\times10^{-6}$, 2.190$\times10^{-6}$, and 5.487$\times10^{-6}$
Jy~(ADU/sec)$^{-1}$ at 1.10, 1.60, and 2.2 $\mu$m (Rieke et al.~1999). The
corresponding magnitude zero points (on the Vega system) were calculated
assuming 1775, 1083 and 668 Jy at 0 mag for 1.10, 1.60, and 2.22 $\mu$m
respectively.  The effective wavelengths of the 110W, 160W and 222M
filters on NIC2 are 1.104, 1.593, and 2.216 $\mu$m (Rieke et al.~1999).

As a check on the flux calibration, we compared our measured fluxes
within a 5\arcsec\ aperture centered on the 1.6$\mu$m and 2.22$\mu$m
peaks in each galaxy with those measured in ground-based imaging by
Carico et al. ~(1988). For the 16 galaxies which overlap between the two
samples, the flux densities of all but one (IR 12112+0305) agree to
better than 10\%.  At 1.1$\mu$m a direct comparison is not possible since
the standard J filter is at 1.25$\mu$m. The rms noise in the final images
is typically 7, 7 and 20 $\mu$Jy (arcsec)$^{-2}$ at 1.1, 1.6, and 2.2
$\mu$m.

\subsection {Point Source Subtraction and Image Artifacts}

In seven of the galaxies imaged here, the images are contaminated by
diffraction rings and spikes due to strong point sources in the galactic
nuclei. These effects are severe in NICMOS images due to near-field
diffraction from the cold-stop. The effects are particularly noticeable at
2.2 $\mu$m but are also seen in some of the 1.6 and 1.1 $\mu$m images. For
the affected images, the nuclear point sources were subtracted
with PSF stars.  Given the variability of the HST PSF caused by telescope
breathing, this removal of the PSF is only approximate. To avoid spurious
PSF artifacts being interpreted as real structure in the images, we
therefore set the surface brightness within a box centered on the point
source to a constant value equal to the average of the flux along the
outside border of the box. This square box had dimensions of 0.66, 0.96
and 1.32\arcsec\ on a side at 1.1, 1.6 and 2.2 $\mu$m respectively. A
point source with the fitted flux was then convolved with a Gaussian of
FWHM = 0.11, 0.16 and 0.22\arcsec\ and added back into the PSF-subtracted
image at the position of the point source in the galaxy. This procedure
was only adopted in the contour diagrams in order to retain information on
the point source strength while at the same time avoiding the display of
areas of the image believed to be particularly unreliable. The procedure
was applied to Figs.~\ref{4contour} and ~\ref{2contour} for NGC 7469, IRAS
08572+3915, IRAS 05189-2524, PKS 1345+12, IRAS 07598+6508, Mrk 1014 and
3C48.  \footnote{Note that no flux measurements were made from images
processed in this manner.} None of the 3-color images displayed in Figure
2 have any PSF subtraction.

Due to the higher resolutions in the two shorter wavelength bands and our
desire to show the original, unconvolved (but calibrated) data for the
3-color images, these images occasionally exhibit `color fringes' in areas
with steep brightness gradients (eg. near a bright point source). The
alternatives (convolving all wavelengths to a common 2.2 $\mu$m resolution
or deconvolving to the 1.1 $\mu$m resolution) would have degraded the
short wavelength resolution or been susceptible to artifacts of the
deconvolution technique. Therefore, the color-fringes appearing in areas
with steep intensity gradients in the 3-color displays should be viewed
with caution and in most cases ignored.

\subsection {Extinction Derivation and Correction}

In most of the galaxies, large spatial variations in the colors of the
near-infrared emission are seen across the images. Aside from the nuclear
point sources (which  may have different intrinsic colors), most of the
large-scale color gradients are probably attributable to variations in the
dust extinction within the galaxy.  This implicitly assumes that outside
the nucleus there is little contribution to the flux from warm dust and
that the galactic disk stellar light in the near infrared is mostly that
of a $\sim 10^8$ yrs stellar population rather than that of the old disk
population (age $\geq 10^9$ yrs).  We thus correct the
 2.2 $\mu$m images for extinction under the assumption that the intrinsic
spectrum of the extended emission is that of an aging starburst population
with the extinction derived from the observed $m_{1.6-2.2}$ $\mu$m color.
For the intrinsic colors, we use the Bruzual \& Charlot (1993, 95) model
starburst population having a Salpeter IMF (see below) over the mass range
0.1-125 \msun\ . For an instantaneous starburst with solar metallicity, we
sample the stellar light in the NICMOS filter set as a function of time.
At ages  5$\times10^{7}$--$5\times10^8$ yrs, the 1.1, 1.6 and 2.2 $\mu$m
colors are fairly constant (see Figure ~\ref{nucflux}) with typical
values being $m_{1.6-2.2}$ $\sim$ 0.35 mag and $m_{1.1-1.6}$ $\sim$ 0.65
mag. Thus, the 1.6-2.2 $\mu$m color excess is given by

$$
E_{1.6-2.2} = -2.5~log({f_{1.6}\over{f_{2.2}}}) +0.175 ,\eqno (1)
$$

To translate the color excess into an extinction we assume that the
extinction is in a foreground screen (i.e. not mixed with the stars) and
use the extinction law derived by Rieke \& Lebofsky (1985) that has been
modified at the shorter wavelengths.  This extinction law translated into
the NICMOS filter bandpasses yields color excesses of $E_{1.6-2.2} \sim
0.079$ and $E_{1.1-1.6} \sim 0.191$ mag for A$_V = 1$ mag.\footnote{Note
that the NICMOS flux calibrations were determined using relatively blue
stars. Thus, in the case of extremely red objects, the broad bandwidth
of the F110W filter may result in an overestimation of their fluxes (e.g.,
by $\sim$0.15 mag for an $A_V \sim 5$).} Therefore,

$$
A_{2.2} = 0.100~{E_{1.6-2.2}\over{0.079}},\eqno (2)
$$
$$
A_{2.2} = -3.15~log({f_{1.6}\over{f_{2.2}}}) +0.222.\eqno (3)
$$
To obtain an image de-extincted at 2.2 $\mu$m, 
$$
f_{2.2}(true) = f_{2.2}(obs)\times~exp({A_{2.2}\over{1.086}}).\eqno (4)
$$
The equivalent relations using the 1.1 and  1.6 $\mu$m bands to estimated
the extinction at 1.6 $\mu$m is
$$
A_{1.6} = -2.30~log({f_{1.1}\over{f_{1.6}}}) -0.105.\eqno (5)
$$
and an image de-extincted at 1.6 $\mu$m is obtained from 
$$
f_{1.6}(true) = f_{1.6}(obs)\times~exp({A_{1.6}\over{1.086}}).\eqno (6)
$$

Equations 3-4 were applied to the 1.6 and 2.2 $\mu$m images, to yield
`de-extincted' 2.2 $\mu$m images for the galaxies. (In a few instances the
2.2 $\mu$m background variations or PSF artifacts severely corrupted the
2.2 $\mu$m image and we used equations 5-6 to obtain `de-extincted' 1.6
$\mu$m images.) To avoid false color gradients in the ratio image used in
Equation (3) (due to the higher resolution at 1.6 $\mu$m), we convolved
both the 1.6 than 2.2 $\mu$m images to a common resolution ($0.2\arcsec$)
before computing the extinction and applying to the 2.2 $\mu$m data (Eq.
4). In Figure \ref{2contour}, the de-extincted 2.2 $\mu$m and 1.1 $\mu$m
(for reference) images are shown for each galaxy. In viewing these
extinction-corrected images, the reader should bear in mind the two
assumptions which are certainly not always correct: that the dust
providing the extinction lies in front of the emission sources (i.e. is
not mixed with the stars) and that the intrinsic (unextincted) color is
uniform and approximated by that of a moderate age starburst population.
The former is probably the most flawed of these assumptions. When the dust
is mixed uniformly with the radiation sources, the observations typically
sample the first few optical depths at each wavelength. For actual $\tau$
in the range 1 to 10 with the dust uniformly mixed with the stars, the
true optical depth is underestimated typically by a factor $\sim 2$ but
for larger $\tau $ the underestimate can be much greater (see
Fig.~\ref{nucflux} and \cite{wit92}). The emission will also appear bluer
than would be expected for the same total extinction. An additional
problem arises from the intrinsically higher angular resolution at 1.6
$\mu$m than at 2.2 $\mu$m. Despite the reservations and assumptions noted
above, these extinction corrected images probably yield a more generally
accurate rendition of the intrinsic 2.2 $\mu$m emission than the observed
2.2 $\mu$m images.  Evidence of this is provided in the case of Arp 220
where the de-extincted image yields peaks which better fit the PA of the
radio nuclei (Scoville et al. 1998).

\section{Morphology}

Morphological characteristics occuring in many of the images of the 24
galaxies include: double nuclei, extended `tidal' tails, point source
nuclei (particularly at 2.2 $\mu$m in the most luminous and distant
objects), bright off-nuclei star clusters, spiral arms (on the scale
of 100 pc to 1 kpc), and high reddening in the nuclei. These features
are summarized in Table 3.  In the following we briefly summarize the
features seen in each object.

\subsection{NGC 4418}
The nucleus of this galaxy exhibits strong extinction along the galactic
disk to within 100 pc of the nucleus which causes the short wavelength
emission (1.1 $\mu$m) to be elongated perpendicular to the plane of the
galaxy at small radii (see Fig.~\ref{4contour}-\ref{2contour}).  The
reddening is most easily seen in the ratio image in Fig.~\ref{4contour}.
Evans et al. (1999a) compare the near-infrared (NICMOS) morphology of the
nuclear region with recently acquired mid-infrared (MIRLIN) images and
discuss the possibility that the bulk of the luminosity of NGC 4418
emanates from a region obscured by a compact ($\sim$ 130 pc) stellar disk.

\subsection{Zw049.057} Zw049.057 appears as a highly inclined disk with a
smooth light distribution at large radii but a bright arm of star
formation $\sim 1\arcsec $ south of the 2.2 $\mu$m peak
(Fig.~\ref{3color}). The reddening and the extinction-corrected 2.2 $\mu$m
emission also peak  $\sim 0.5\arcsec $ south of the 2.2 $\mu$m emission
peak (Fig.~\ref{4contour} and~\ref{2contour}). Zw049.057 exhibits a linear
'shadow' feature extending radially from the nucleus along the minor axis
at $PA \simeq -50\deg$ (see Fig.~\ref{3color}). The feature is perhaps due
to an absorbing cloud in the nucleus blocking radiation along the shadow
line on the minor axis. This interpretation would imply that much of the
light seen along the minor axis of the galaxy is scattered light from the
nucleus. Alternatively, if this is a dust absorption lane, it must extend
{\it{linearly}} over 500 pc in radius and a high mass is required in order
to produce the absorption over the extent of the lane.

\subsection{NGC 6090}
In the primary (NE) galaxy of NGC 6090, two distorted spiral arms are
seen, both of which are delineated by a number of bright clusters. The
companion, 6\arcsec (3.4 kpc) to the SW could almost be an extension of
one of the primary's spiral arms; however, the bright 2.2 $\mu$m point
source at the one end of the secondary together with a radio continuum
source argue for this being a less massive galaxy which is merging with
the NE galaxy. The 2.2 $\mu$m extinction-corrected contour map
(Fig.~\ref{2contour}) strongly favors this interpretation since the SW
galaxy clearly appears very substantial and has only a low level bridge to
the spiral pattern of the NE galaxy. The area between the two galaxies
contains a massive concentration of ISM as evidenced by the very red
colors of the clusters (No. 6 and 8 in Table 7) on the SW edge of the
primary galaxy and the fact that the mm-wave CO(1-0) emission peaks in
this overlap region (\cite{bry99}). The morphology of NGC 6090 suggests an
extended starburst triggered by the galaxy-galaxy interaction.  The large
number of luminous clusters seen along the side of the NE galaxy closest
to the secondary galaxy suggests that the starbursts are triggered
hydrodynamically (eg. cloud-cloud collisions or shocks from a high
pressure, intercloud medium) rather than by large-scale gravitational
force gradients. (Tidal effects should be equal on the near and far sides
contrary to the observed asymmetry.) Dinshaw et al. (1999) present a more
detailed description of these data, and conclude, in part, that the radio
emission from NGC 6090SW is not coincident with the brightest
near-infrared ``knot'', and that this knot may actually be a forground
star.

\subsection{NGC 2623}
 NGC 2623 has a highly reddened nucleus with a possible short tidal
feature or spiral arm at $\sim 1\arcsec$ radius to the SW (seen best
in the 1.1  $\mu$m image). In the optical, prominent tails are seen
suggesting that this system underwent a significant merging event in
the past (Toomre 1977, Joseph \& Wright 1985).  However, at 1.6 and
2.2 $\mu$m, the light profiles are smooth, following approximately an
$r^{1/4}$ law (\cite{wri90} see below). The latter morphology suggests
that the galaxies which may have merged have coalesced into a common
nucleus. A weak VLBI radio core is seen in the nucleus of NGC 2623
(cf. \cite{lon93}). In CO(1-0) the emission complex is $\sim 1.6$\arcsec\
in diameter with kinematic major axis E-W (\cite{bry99}), similar to
the reddening distribution shown in Fig.~\ref{4contour}.

\subsection{IC 883}
IC 883 (Arp 193) appears as a highly inclined disk with the reddening
increasing to the NW within the disk. The peak in the extinction-corrected
2.2 $\mu$m light distribution coincides with the 2.2 $\mu$m flux peak
but the centroid is clearly displaced to the northwest.  The reddening
distribution shown in Fig.~\ref{4contour} is similar to the CO(1-0)
emission which has a size $4.1\arcsec\times2.2\arcsec$ elongated along the
plane of the galaxy and with kinematic major axis in the same direction
(\cite{bry99}). (The increase in apparent 2.2/1.1 $\mu$m color ratio on
the NE of IC 883 is small and occuring at low flux flux levels; it may
be due to flat-fielding errors.) A number a bright clusters are seen
above and below the disk out to 5\arcsec\ radius. Their high luminosity
suggests that they are young ($\leq 10^9$ yrs; see below), implying that
the galaxy may have undergone a collision in the past with a burst of
star and cluster formation in spherical region before the ISM settled
into its present disk-like configuration. The near-infrared morphology of
IC 883 is very similar to that of M82 although the luminosity is scaled
up by over an order of magnitude.

\subsection{NGC 7469}
The near-infrared emission from NGC 7469 is dominated by the bright,
point-like Seyfert nucleus point-source.  However, the bright inner disk
of the galaxy has been seen in much earlier optical (De Robertis \& Pogge
1986, Wilson et al. 1986) and near-infrared (\cite{maz94}) imaging, and is
discussed extensively in Genzel et al.  (1995). A short spiral arm-like
feature in the NW are clearly seen in the point-source subtracted images
(Fig.~\ref{4contour}).  (The bright linear SSE--NNW feature is a PSF
artifact which we were unable to remove in the 2.2 $\mu$m image.) In the
inner disk, there is a ring of star formation at $\sim 1$\arcsec\ radius,
corresponding to 500 pc. The structure within the ring is similar in all
three bands and is well outside the area in which residual PSF should
introduce structure. This starburst disk is very similar to that seen in
NGC 1068; in NGC 7469 there is no evidence of a bar like that seen in NGC
1068.  The secondary companion to NGC 7469, IC 5283 is 80\arcsec\ (26 kpc)
away and therefore well outside the field of our images. The data for this
galaxy will be discussed more thoroughly in Thompson et al. (1999).

\subsection{VV 114}
The two galaxies in VV 114 are sufficiently extended that a mosaic of two
images was required to cover them both using the NIC2 camera.  These
galaxies present a remarkable contrast from the visual to near-infrared
(cf. \cite{kno94}; \cite{doy95}) -- the eastern galaxy which is very
insignificant in the visual becomes the dominant source in the
near-infrared in terms of surface brightness.  The western galaxy has a
great number of luminous star forming regions in an arm running between
the two galaxies and to its south. This positioning of the young clusters
in the overlap region of the two galaxies is similar to that seen in NGC
6090. In the mm-wave CO line (Yun, Scoville \& Knop 1995) and 850 $\mu$m
continuum (Frayer et al. 1999), the major concentration of emission
actually lies between the two galaxies .  In the eastern galaxy, there are
two 2 $\mu$m peaks, the brightest being that to the SW where the reddening
is also greatest. A more detailed discussion of this galaxy is provided in
Scoville et al. (1999a).

\subsection{NGC 6240} The double nuclei in NGC 6240 are separated by
1.6\arcsec\ N-S (0.8 kpc) and the southern nucleus is relatively brighter
at long wavelengths.  The reddening peaks to the north and slightly east
of the southern nucleus.  In the radio continuum, there are also two
nuclei but their separation is only 1.4\arcsec (\cite{crl90}). The mm-wave
CO(2-1) emission peak is located between these nuclei (Tacconi et al.~1999,
Bryant \& Scoville 1999). In addition, the near-infrared CO-bandhead
velocity dispersion of giant stars exhibits an increase between the nuclei
(Tacconi et al. 1997), all of which indicate a significant mass
concentration (perhaps largely interstellar gas) between the nuclei.

\subsection{VIIZw031}
Ground-based optical images have been used to suggest that VIIZw031 might
be an elliptical galaxy (eg. Djorgovski et al. 1990; \cite{san96}) ;
however, the NICMOS images clearly resolve extremely bright, asymmetric
spiral arms in the nucleus.  Numerous clusters are seen in the galactic
disk and the reddening peaks to the east of the nucleus. While the NICMOS
data clearly imply that this is a spiral system and strengthens the view
that the progenitors of LIGs and ULIGs are composed of at least one spiral
galaxy, it is not obvious what triggered the activity occuring in
VIIZw031. There is no evidence to date of a nearby interacting companion
galaxy (pre-merger) or tidal tail remnants (post-merger).

\subsection{IRAS 15250+3609} In IRAS 15250+3609 there is one dominant
nucleus and a much less bright (possible) nucleus 0.7\arcsec\ SE. It is
not clear if this second source is indeed a nucleus or an anomalously
bright cluster but its colors are very red like most of the nuclei in our
sample and it is approximately a factor of ten more luminous than any
other cluster in this system. The emission associated with the primary
nucleus is also extended in the direction of the secondary source. In
addition, the nuclear region is surrounded by several bright star
clusters.

\subsection{UGC 5101}
UGC 5101 (IRAS 09320+6134) shows a single nucleus with surrounding spiral
isophotes that rotate in PA as a function of radius. Its nucleus is
unresolved at 2.2 $\mu$m and extremely red.  In the color-color diagrams
(Fig.~\ref{nucflux}-\ref{nucflux}), the nucleus is anomalous in being the
only one of the cold ULIGs with colors similar to warm ULIGs
(possibly indicating an AGN source). In the optical, an extended edge-on
tidal tail is seen like that in Mrk 273 (see {\cite{san88a}}, Surace et
al. 1999b), while a second tail loops around the nucleus in a nearly
complete ring.  A number of clusters can be seen in the northern arm and
the reddening increases just to the north of the nucleus. Due the bright
galactic background near the nucleus, we were not able to accurately fit
and subtract a PSF from the 2.2 $\mu$m image and the SW--NE stripe is
a residual PSF artifact.

\subsection{IRAS 10565+2448}
The primary galaxy in IRAS 10565+2448 is much more luminous than the
companion galaxy, located near the edge of our images 8\arcsec\ to the SE.
Nevertheles, it appears that the two are interacting given the bridge
between them (see Fig.~\ref{3color}) and their angular separation which
corresponds to only 6.7 kpc. A third galaxy and tidal tail is seen to the
NE of the primary, but out of the field of view of our NICMOS images
({\cite{mur96}}).

\subsection{IRAS 08572+3915}
The two nuclei in this system are separated by 5\arcsec, corresponding to
5.6 kpc.  The northern nucleus is unresolved in all three bands and an
extremely bright cluster is seen to the SE of the nucleus
(Fig.~\ref{3color}). The low level common envelope of the system which is
best seen in the contour images (Fig.~\ref{4contour}-\ref{2contour})
bridges the region between the galaxies at 1.1 and 1.6 $\mu$m and extends
well to the east of the southern galaxy. (The disappearance of the
envelope in the 2.2 $\mu$m image in Fig.~\ref{4contour} is probably due to
the higher background and lower sensitivty at 2.2 $\mu$m.)

\subsection{IRAS 05189-2524} This galaxy shows a single unresolved
nucleus in all three bands. In the PSF-subtracted images used in
Fig.~\ref{4contour}, low level emission is seen to at least 3\arcsec\
radius. The nuclear source is extremely red (see Fig.~\ref{4contour}).
(The sharper extensions E-W and to the N in the 2.2 $\mu$m image of
Fig. ~\ref{4contour}-~\ref{2contour} are probably due to incomplete
PSF removal.) At optical wavelengths (Surace \& Sanders 1998,1999),
this galaxy has a nucleus which appears to be bisected by a dust
lane; the longer wavelength NICMOS data clearly penetrates this dust.
The optical images also exhibit a ``plateau'' of extended blue star
formation surrounding the nucleus which may correspond to the extended
light seen in the NICMOS data, as well as several extended tidal loops.

\subsection{IRAS 22491-1808}
This spectacular system has two nuclei separated by
1.6\arcsec\ (2.4 kpc), which were first observed by Carico et al. (1990b).
Many extremely luminous clusters and two high surface brightness
tidal tails are seen extending to E and NW. The western nucleus is
unresolved in all three bands. Optical imaging reveals so many luminous
star clusters as to prevent identification of the actual nuclei;
near-infrared data are required to locate them (Surace 1998, Surace et al.
1999b). The optical images also reveal that the full extent of the tails,
which terminate in complex loops. Many of the clusters seen by NICMOS are
embedded in these tails, particularly the one to the NW.

\subsection{Mrk 273}
The northern nucleus is both bluer and more spatially extended than the
redder, unresolved southern nucleus.
Both nuclei are much redder than the surrounding galaxy.
The tidal tails may be seen in both the 1.1 $\mu$m emission and in
obscuration (2.2/1.1 $\mu$m) extending well to the north and south of the
nuclei. In addition, several bright clusters are seen to the south and
north of the northern nucleus.  Optical imaging shows the presence of a
northern tidal loop as well as an edge- on tidal feature to the south
(Sanders et al. 1988a). Additional optical and UV imaging shows the
presence of young star formation in the northern nucleus and in a region
directly to its west (Surace 1998). Knapen et al. (1997) discuss the
relation of their adaptive optics near-infrared data to their radio data.
Surprisingly, they find that the northern nucleus is the strongest radio
source and associate it with the active nucleus which is presumably the
source of the known Sy 2 emission (Sanders et al. 1988a), while they find
the dominant infrared peak (the SW nucleus) to be the location of a
starburst, which is the opposite of what is implied based on the NICMOS
morphology. A near-infrared source SE of the northern nucleus is also
shown which had been known previously (Condon et al. 1991). While Knapen
et al. present this as a background object, the NICMOS data shows it is
spatially coincident with a blue compact object identified as a star
cluster.

\subsection{Arp 220}
This merging system contains two nuclei in both the radio continuum and
the near-infrared separated by 0.95\arcsec (350 pc projected separation).
A detailed discussion of the NICMOS data on this galaxy is given
in Scoville et al. (1998).  In the data, extremely high reddening
gradients are seen to the south of the brighter western nucleus
and to the south of the eastern peak seen in the 1.1 $\mu$m images
(Fig.~\ref{4contour}).  In fact, the extinction is so high on the east
that in the extinction-corrected image (Fig.~\ref{2contour}), the eastern
peak which we identify with the eastern nucleus lies well to the south
of the 1.1 $\mu$m peak, between this peak and a weak third peak seen in
the 2.2 $\mu$m image. The strong obscuration to the south of the western
peak can be interpreted as an inclined dust disk ($i \simeq 30\deg$)
embedded in the nuclear star cluster (Scoville et al. 1998). The light
which escapes the cluster can appear crescent-shaped if the disk is of
size comparable to (or smaller than) the cluster and embedded within
the cluster.

\subsection{PKS 1345+12} The western nucleus in PKS 1345+12 is unresolved
in all three bands and much redder than the eastern nucleus (see
Fig.~\ref{4contour}).  The eastern nucleus is clearly resolved and
extended. The common envelope for the two nuclei is extended in the
east-western direction, as well as to the south.  Evans et al. (1999b)
compare the near-infrared (NICMOS) data with radio data and recently
acquired CO($1\to0$) interferometry to show that the molecular gas and
radio jets of PKS 1345+12 are associated with the redder nucleus, and thus
that the molecular gas is the likely source of fuel for the imbedded,
radio-loud AGN.

\subsection{IRAS 12112+0305}
The two well-separated nuclei in this system were first observed by Carico
et al. (1990b), and are apparently connected by a bridge of emission,
and a tidal tail 4\arcsec\ SW of the southern nucleus. The northern
nucleus is crescent-shaped (like that in Arp 220) and this may indicate
the presence of an embedded opaque dust disk (see above). The southern
nucleus becomes increasingly point-like at the longer wavelengths
and both nuclei are much redder than the surrounding galaxy (see
Fig.~\ref{4contour}). Optical imaging reveals a northern counter-tail
(Surace et al. 1998b) while ultraviolet imaging reveals the presence of
significant obscuration along the line of sight to the southern nucleus,
as well as young star-forming clusters embedded in the southern tail
(Surace \& Sanders 1999c).

\subsection{IRAS 14348-1447} This system has two well-separated but
clearly interacting spiral galaxies -- a curved tidal tail with embedded
clusters can be seen to the NE of the northern galaxy and a less extended,
fainter tail is seen to the SW of the southern galaxy.  The southern
nucleus has a ring of star clusters to its northwest and southwest; these
are best seen in optical images (Surace 1998) and do not readily appear in
the NICMOS data, which is indicative of their young age.

\subsection{IRAS 17208-0018} IRAS 17208-0018 is the most luminous galaxy
in our sample that shows no direct evidence of an AGN -- the optical
emission line ratios are HII-like, the nucleus in the near-infrared images
is extended in all bands, and the inner disk at $R\leq 1$ kpc shows
numerous extremely luminous clusters. The same region has very strong
reddening gradients -- it is quite likely that even at 2.2 $\mu$m, dust is
still masking the nucleus. The outer disk of the galaxy appears very
disturbed (best seen in the 3-color image, Fig.~\ref{3color}).

\subsection{IRAS 07598+6508} The broad-band emission in all three filters is
dominated by the point-source nucleus but low level emission can also be
seen out to $\sim 2$\arcsec\ radius. Optical {\it HST} images reveal the
presence of luminous blue star clusters to the south and east (Boyce et al.
1996), and it is unclear if the extended emission is due to these clusters
or to the underlying host galaxy.

\subsection{Mrk 1014}
Although dominated by the QSO nucleus, point-source subtraction clearly
shows twisting spiral isophotes within the central 4 kpc, indicating either
a starburst spiral disk or tidal debris (see Fig. ~\ref{4contour}). Similar
features are seen optically in the inner nuclear regions (Surace et al. 1998).
Wide-field deep optical imaging reveals a tidal arm extending to the NE over
a distance of 60 kpc (MacKenty \& Stockton 1984) which has many embedded
star clusters (Surace 1998), which are too blue to see here.

\subsection{3C48} In 3C48, the emission in all three bands is dominated
by the unresolved quasar nucleus; however after point-source subtraction
(see Fig.~\ref{4contour}) extended emission can be seen (particularly at
1.6 $\mu$m) to the NE and S of the nucleus. These extensions correspond
to those seen in the optical and K-band by Stockton \& Ridgway (1991);
they identify the NE source and the nucleus of the galaxy merging with the
QSO host galaxy. (3C48 was not included in Fig.~\ref{2contour} since the
ratio image is severely contaminated by the PSF artifacts at 2 $\mu$m).

\section{Nuclear Bars and Spirals}

It is often suggested that nuclear bars or spiral structure made aid in
the loss of angular momentum from the gas and hence lead to high rates of
radial accretion to feed nuclear activity (e.g. Schlosman, Frank \&
Begelman 1989). The high resolution NICMOS images of our galaxy sample
provide some constraints on such structures. In Table 3, 10 of the
galaxies are listed as having spiral-like structures in the inner kpc and
in most cases these spiral arms continue into $\leq$ 100 pc radius. In no
cases did we find evidence of an obvious nuclear bar on similar scales;
however, it must be recognized that most of these systems are sufficiently
disordered due to variable extinction and starburst activity that an
underlying bar in the older stellar population would probably be difficult
to detect even if it were there. This is in agreement with Regan \&
Mulchaey (1999) who used optical and near infrared HST imaging for a
sample of nearby AGN to look for small scale nuclear bars in reddening
distributions. Only 3 out of 12 Seyfert galaxies in their sample exhibit
nuclear bars.

\section {Flux Measurements}

Measurements of the magnitudes at 1.1, 1.6 and 2.2 $\mu$m were made for
each galaxy in 1.1, 5 and 11.4\arcsec\ diameter circular apertures centered
on the nucleus using the IPAC routine SKYVIEW (note that the 1.1$\arcsec$
aperture was used in order to include the first Airy ring of the NICMOS
PSF). The resultant magnitudes are listed in Tables 3 and 4. For six of the
galaxies with well resolved double nuclei, both galactic nuclei were measured.
In addition to the aperture measurements, the magnitudes of the nuclei were
determined by subtracting the contribution from the underlying stellar light
(i.e., the adjacent pixels) from the measured 1.1\arcsec\ nuclear apertures.
The results are listed in Table 5.

All compact (cluster) sources outside the nuclei were measured in a
0.53$\arcsec$ diameter aperture using Source Extractor (Bertin \& Arnouts
1996), and the results are tabulated in Table 7 for all sources with a
signal-to-noise ratio greater than 3.0 in all three bands. As was the case
with the nuclear magnitudes, the local background of every cluster was
subtracted based on a sampling of the adjacent emission.  

\subsection{Color-Color Diagrams for Galaxies}

In Figures~\ref{nucflux}-\ref{2kpcflux}, the $m_{1.1-1.6}$ and
$m_{1.6-2.2}$ colors of the galaxy sample are plotted for the nuclei with
symbols denoting the infrared luminosity and `warm/cool' classification
(Table 5), with symbols denoting the optical emission-line classification,
and for a fixed 2 kpc-diameter aperture (Table 6) with symbols denoting
the infrared luminosity and `warm/cool' classification.  Also shown in
the figures are the expected colors for an instantaneous starburst model
(Salpeter IMF with masses ranging from 0.1 -- 125 \msun and solar
metallicity) as it ages (\cite{bru93}). In this model, the colors change
rapidly during the first $10^7$ yr but after that they are relatively
constant out to $5\times10^8$ yr. The mean color of optical bright PG QSOs
(\cite{san88b}) is shown by the cross and the dotted lines shows the
effect of additional emission by hot dust with increasing contributions
at  2.2 $\mu$m. It is important to note that the near-infrared SED of the
PG QSO probably already has contributions due to hot dust and these curves
represent an arbitrary increase in the relative dust contributions, not a
new generic component (\cite{bar87}). The color of free-free emission is
shown and the effects of differing amounts of extinction is indicated by
the reddening vector in the lower left of the figures. The reddening
vector was calculated from the extinction curve of Rieke \& Lebofsky
(1985) and Whittet et al. (1998) assuming a foreground dust screen. For
comparison, the reddening path for extinction by dust mixed uniformly with
the stars is also shown (the curved track).  Both the model fluxes and the
extinction curve were convolved with the NICMOS filter bandpasses. The
models were also redshifted to z$= 0.05$, corresponding to a typical
redshift of the galaxies observed here.

The salient features of the color-color diagrams are :

1) Virtually all of the
galaxies are redder in both $m_{1.1-1.6}$ and $m_{1.6-2.2}$ than the
unreddened starburst colors (i.e. they lie well above and to the right of
the starburst evolutionary track, see Fig.~\ref{nucflux}). Their colors
clearly require either extincted starlight and/or an AGN energy
source.  This general trend has been observed in ground-based near-infrared
data as well (e.g. \cite{san88a}; \cite{car90a}; \cite{maz92}).

2) Both nuclei of Arp 220 and the nucleus of IR 17208-0018 are observed
to have the most extreme $m_{1.1-1.6}$ colors of all of the galaxies in
the sample. Thus, if the near-infrared light is associated with the source
of the bolometric luminosity, then the nuclear power sources in these two
galaxies are well buried even at near-infrared wavelengths.

3) The warm ULIGs (IRAS 05189-2524, IRAS 07598+6508, PKS 1345+12, Mrk 1014
IRAS 08572+3915, and 3C 48) are all much redder in $m_{1.6-2.2}$ than
$m_{1.1-1.6}$ relative to a typical cold ULIG or a PG QSO (see
Fig..~\ref{nucflux}), consistent with similar analyses
done with ground-based observations (e.g. \cite{sur99a}) . These
colors for the warm ULIGs are hard to account for by reddened starlight.
The very red $m_{1.6-2.2}$ color is probably due to substantial
contribution at 2.2 $\mu$m by warm dust emission (at 600-1000 K).

4) On average, the nuclear colors of the cold ULIGs are redder than those
of the less-luminous cold LIGs (see Fig..~\ref{nucflux}). Of the three
ULIGs that contain at least one nucleus similar in color to the LIGs (IR
08572+3915S, PKS 1345+12E, and IR 22491-1808W), this nucleus is
significantly bluer than the other.  The two LIGs observed to have
extremely red colors (comparable to Arp 220) are IC 883 and NGC 2623.

5) The nuclei in VV114E and UGC 5101 have near-infrared colors which place
them in the area of the color-color diagram occupied by warm ULIGs despite
the fact that their mid-infrared colors are cold (note that the extremely
red colors of VV114E have been previously discussed by \cite{kno94}).  Their
nuclear colors (Fig..~\ref{nucflux}) could be explained by the dust mixed with
stars model, but it would require $>>50$ magnitudes of visual extinction.
The underlying galaxy-subtracted nuclear colors of Arp 220E, Mrk 273S, and
IR 17208-0018 (Table 5; Fig..~\ref{nucflux}) also place these galaxies in
the warm galaxy section of the plot, but require a 60--70\% contribution
to the 2.2 $\mu$m light by 1000K dust.  Mid-infrared spectroscopy with ISO
revealed high ionization state emission lines in Mrk 273 (\cite{gen98}),
but it likely that these lines emanate from the northern nucleus.  These may
be examples of galaxies having substantial starbursts together with AGNs.

6) On average, the ULIGs that are optically classified as seyferts are
redder (in $m_{1.6-2.2}$) than the HII region-like galaxies
(Fig..~\ref{diagnflux}. With the exception of IR 08572+3915S, the LINERs
appear to be dispersed over the region redward of $m_{1.1-1.6} = 1.5$ and
$m_{1.6-2.2} = 0.5$.

7) In the 2 kpc-diameter aperture measurement (Fig.~\ref{2kpcflux}), the
cold ULIGs and the LIGs move back toward the starburst models with a few
magnitudes of visual extinction.  The outer ULIG regions are thus
consistent with simple star formation models with reddening corresponding
to {\it A}$_{\rm V}$ of a few mag., assuming a foreground screen of dust.

\subsection{Color-Color Diagram for Clusters}

In Fig.~\ref{cluflux} the $m_{1.1-1.6}$ and $m_{1.6-2.2}$ colors are
shown for all of the measured clusters which have $SNR \geq 5$
in all three bands.  While there is a large spread in both colors,
most of the clusters have $m_{1.1-1.6} < 1.5$ mag and 
$m_{1.6-2.2} < 1.0$ mag.  The majority are consistent with 
young star clusters (ages between 5 and 300 Myr) which are
reddened by up to 3 magnitudes of visual extinction.
The majority of the cluster sources have colors
implying much lower reddening than the galactic nuclei (compare
Figs.~\ref{nucflux} and \ref{cluflux}). Some of this is due to the fact
that the clusters are most easily detected outside the nucleus where the
dust extinction is less.  The cluster extinction is also likely to be
better-characterized by a foreground-screen model since it is unlikely there
is much dust {\it{inside}} the clusters (cf.\cite{whi95}). 

The observed magnitudes of the clusters can be used to crudely estimate
their bolometric luminosities and masses with some simplifying
assumptions.  First, the instantaneous starburst models (described
earlier) exhibit steeply rising near-infrared fluxes up to $10^7$ yrs, at
which point the 1-2.2 $\mu$m fluxes change relatively slowly out to $10^9$
yrs. This is due to the dominance in the near-infrared of red supergiants
from the starburst.  Secondly, most of the cluster formation was probably
triggered {\it {dynamically}} by galactic interaction and merging over a
time $\geq 3\times10^7$ yrs. The combination of these two
{\it{reasonable}} assumptions, implies that most of the observed clusters
are likely to be of age $10^7$--$10^8$ yrs -- a period during which the
colors are relatively constant and the bolometric corrections to convert
from 1.6 $\mu$m fluxes to bolometric luminosity are also relatively
constant.  For For ages of $3\times10^6, 10^7, 10^8$ and $10^9$ yrs, the
1.6 $\mu$m bolometric corrections are  -3.5, +0.45, +0.67 and +1.7 mag
respectively (Bruzual \& Charlot 1993).  For the estimates below, we
simply adopt a fixed bolometric correction for the 1.6 $\mu$m band (i.e.
M$_{bol}$=M$_{1.6}$+0.6 mag), and assume no reddening. The absolute
M$_{\bf1.6}$ are in the range -14.89 to -19.01 mag for the first-ranked
clusters (Table 7). The derived bolometric luminosities of the brightest
clusters range from 4$\times 10^7$ to 2$\times 10^9$\lsun\ with a median
value of 4$\times 10^8$$L_\odot$  .

The BC95 (Bruzual \& Charlot 1993: an updated version called BC95 is used
here, private communication) model for an instantaneous starburst with a
Salpeter IMF between 0.1 and 125 \msun\ can then be used to estimate the
required cluster mass from the 1.6 $\mu$m absolute magnitudes.  At a
typical age of 5$\times 10^7$ yrs, the luminosity to mass ratio of these
models is 15 \lsun\ / \msun\, yielding mass of the first-ranked clusters
in the range  3$\times 10^6$ to 1$\times 10^8$\msun\ with a median value
of 3$\times 10^7$ M$_\odot$.  If the clusters were older than assumed above,
the implied masses are of course greater. And for an IMF tuncated on the
low mass end at 2.5\msun\ (instead of 0.1\msun\ ), the dervied masses are
typically a factor of 3-4 less.  For comparison, Galactic globular
clusters have typical masses $\sim 10^5$\msun (\cite{van95}) and even if
the clusters seen here have a truncated IMFs, the inferred masses are well
in excess of those expected for globular clusters. The clusters seen in
the infrared luminous galaxies must therefore be super-massive compared to
known globular clusters and/or be unresolved associations of many globular
clusters.  The very high masses implied by the assumption of a standard IMF
extending to 0.1 \msun\ strongly suggest that the IMF cutoff at a higher
mass. This is similar to the conclusion derived from optical imaging by
Surace et al. (1998) based on the measured sizes and brightnesses of the
clusters  in HST optical images.

\section {Radial Distributions}

The degree of nuclear concentration of the light as a function of
wavelength, luminosity and galaxy type (i.e. optical spectral class or IR
warm vs cold colors) can provide important clues to the
luminosity source and evolutionary state of the galaxies.  In this
section, we first discuss the radial surface brightness distributions
before quantifying the central concentrations in terms of the half-light
radii.

\subsection {Surface Brightness}

For each of the galaxy images, the radial surface brightness profile was
computed from the mean brightness of pixels within each radial bin. In
Figs.~\ref{radial_kpc}, the surface brightness (Jy arcsec$^{-2}$) is shown as
a function of angular (top scale) and linear radius (bottom scale) for each
galaxy. In view of the irregular morphologies of most of these galaxies, we
did not fit elliptical isophotes to the image data but instead just adopted
the apparent (projected) radius for these plots. The dynamic range from peak
to the level of undetectable emission is typically three orders of magnitude.
In VV114 where there are two well-separated galaxies of comparable magnitude,
a radial profile was done separately for each galaxy; in the other cases
only one profile was measured from the 2.2 $\mu$m peak position.  In systems
with two galaxies, the radial profiles show a second peak simply due to
the secondary galaxy (eg. IRAS 12112+0305 and IRAS 14348-1447).

Comparison of the light profiles in the three bands clearly shows that in
virtually all cases (except VV114W) the 2.2 $\mu$m flux (solid line in
Figs.~\ref{radial_kpc}) is more centrally peaked
at the nucleus than that at 1.1 and 1.6 $\mu$m (dotted and dashed
lines). Throughout most of the galaxies, the absolute value of the
surface brightness in Jy arcsec$^{-2}$ is also higher at the longer
wavelengths. The relative flux variations imply the stellar light must be
highly reddened by dust.  Moreover, both the dust and the youthful
population must be concentrated at smaller radii. For the starburst
population described in \S 2.3, the colors are $m_{1.6-2.2}$ $\sim$ 0.35
mag and $m_{1.1-1.6}$ $\sim$ 0.65 mag after 5$\times10^{7}$ yrs which
corresponds to $f_{1.6}/f_{2.2} \sim$ 1.18 and $f_{1.1}/f_{1.6}$ $\sim$
0.97. Since these colors are approximately the reddest obtained in the
starburst models, the observed colors, with long wavelength fluxes
exceeding the shorter wavelength, necessitate substantial reddening and
cannot be attributed solely to a youthful stellar population.

In NGC 4418 and VIIZw031, the outer regions of the galaxies have
apparently lower surface brightness at 2.2 $\mu$m than in the 1.1 and 1.6
$\mu$m bands; however, this occurs at low flux levels where variations in
the 2.2 $\mu$m background may be responsible. In VV114W the higher flux at
shorter wavelengths occurs at higher flux levels and clearly real.

\subsection { ${\bf R^{1/4} }$ and Exponential Disk Radial Profiles}

Since it has been postulated that ULIGs may evolve into elliptical
galaxies once the starburst subsides and the gas is either used up or
expelled in a wind (Sanders et al. 1988a; Heckman, Armus \& Miley 1990)
the stellar surface brightness profiles of ULIGs might shed light upon
their evolutionary state.  To measure the stellar light distribution in
dusty systems, it is wise to go to the longest possible wavelengths, while
still avoiding significant contamination from warm dust.  Thus the near
infrared is the ideal wavelength at which to measure the distribution of
stars in ULIGs and compare the average profiles to models fitting broad
classes of elliptical or spiral Hubble types.  Schweizer (1982) first
showed that elliptical-like light profiles could be seen in a merging
system, in this case NGC 7252 in the V-band, and thus provide strong
evidence for relaxation of the stellar population and eventual evolution
into an elliptical galaxy.  Among the first to perform this experiment for
LIGs and ULIGs were Wright et al. (1990), who noted that the
near--infrared K-band surface brightness profiles of Arp 220 and NGC 2623
could be well fit with a de Vaucouleurs r$^{1/4}$ law (i.e. log$\Sigma
\sim$r$^{1/4}$) over reasonably large radii ($\sim0.5-4$ Kpc for Arp 220
and $\sim0.8-4$ Kpc for NGC 2623).  It is important to note that both Arp
220 and NGC 2623 could also be well fit by exponential (spiral-like)
surface brightness profiles, but only over a smaller range in radii
($\sim2-4.5$ Kpc for Arp 220 and NGC 2623).

The superior resolving power and sensitivity to small--scale features that
is possible with NICMOS makes an exploration of the radial surface
brightness profiles of the galaxies in our sample worthwhile.  We show the
$1.6\mu$m logarithmic surface brightness profile for each galaxy as a
linear function of radius in Fig.~\ref{radial_kpc}.  Here, a straight line
fit would suggest a spiral-like stellar profile.  Similarly, we show the
$1.6\mu$m logarithmic surface brightness for each system as a function of
r$^{1/4}$ in Fig.~\ref{radial_r1_4}. An r$^{1/4}$ law profile will appear
as a straight line in this figure.  To evaluate quantitatively the
r$^{1/4}$-law and exponential-disk models, minimum $\chi^{2}$ fitting of
both models to the observational data was done and Table 8 summarizes the
results. The radial range for the fitting was between 0.22\arcsec\ radius
and the outer limit of detectable emission (typically a factor of 30 in
radius). All model fitting was performed on the 1.6 $\mu$m data (as
opposed to the $2.2\mu$m data) in order to take advantage of the greater
sensitivity of NICMOS at $1.6\mu$m. (Because the $2.2\mu$m fluxes are less
reliable in the faint, outer regions of the galaxies, we also did not use
the extinction-corrected surface brightness.)

In 9 of 24 galaxies, the light profiles at 1.6 $\mu$m are fit better, (in
the sense that the ratio of the $\chi^{2}$ from the two fits is greater
than three) by an r$^{1/4}$-law than by an exponential disk profile. These
galaxies are NGC 4418, Zw049.057, NGC 2623, IC883, NGC 6240, UGC 5101,
IRAS 10565+2448, Arp 220, IRAS 14348-1447. Several of these galaxies have
been recognized in previous work as exhibiting an r$^{1/4}$ law : NGC 2623
(\cite{wri90}; \cite{sta91}), IC883 (\cite{sta91}), and Arp 220
(\cite{wri90}) and the scale lengths derived here (see Table 8) are
similar to those derived previously by these authors. In the case of NGC
2623, the exponential disk form becomes acceptable if the nucleus is
excluded.  It is important to note, that the majority of the systems studied
here (13 of 24) could be fit equally well by either an r$^{1/4}$-law or an
exponential profile over the radii we are exploring with NICMOS. In only
in one case, VV114E, was an exponential fit significantly better than
an r$^{1/4}$-law fit. We also find that 8 of the 9 galaxies fit well by
the r$^{1/4}$-law have cool IRAS colors (or equivalently have HII-like
optical emission lines) while only one is warm. This correlation might
arise since the cool ULIGs have more extended far infrared emission from
an extended starburst population. The young stars, formed in a dynamically
relaxed merger, eventually evolve to become bright in the near-infrared --
possibly resulting in an  r$^{1/4}$-law near-infrared light distribution.

Our admittedly small statistics seem to indicate for those ULIGs and LIGs
where a clear preference is shown for one type of fit over the other, this
is usually an r$^{1/4}$-law profile.  Since our measurements cover a large
range in radii for each source, and extend well beyond a typical spiral
bulge radius in nearly all cases, this result is not simply stating the
obvious fact that merging spiral galaxies have elliptical-like bulges.
Instead, it suggests that the stellar population whose light dominates the
inner $5-10$ Kpc in these galaxies appears to be better approximated by a
spheroidal as opposed to a disk-like orbital configuration.  If the
near--infrared light is dominated by young stars such as red supergiants,
these stars must have formed during the merger and have already assumed
elliptical-like orbits.  Whether these systems end up forming giant
ellipticals will depend on a number of factors -- most importantly, the
overall mass density of stars in the central regions and the quantity of
ISM left over in a cold disk after the merging is complete. Kormendy \&
Sanders (1992) have pointed out that in some of the ultra-luminous sytems
(eg. Arp 220), the central mass density is in fact similar to the of
elliptical galaxy cores if the massive ISM component is included -- the
presumption is then that an elliptical galaxy could be the end product if
a significant fraction of the ISM is converted into stars. However, these
are extreme examples and it seems more likely that the lower luminousity
infrared, merging galaxies will end up as spirals with massive central
bulges.

\section {Curve of Growth and Half-Light Radii}

The change in color and/or the degree of compactness in the light
distribution of a galaxy can provide important clues as to the dominant
energy source.  An active galaxy (Seyfert or quasar) can emit a
significant fraction of its energy at all wavelengths on very small
scales (tens of pc or smaller), whereas this is impossible for a
starburst.  The more luminous a system is, the harder it is to pack the
requisite number of young stars and supernovae into a small enough volume
without disrupting the starburst completely via the action of stellar and
supernovae-driven winds.  Although star formation in luminous infrared
galaxies is believed to occur over relatively small (as compared to an
entire galaxy) scales, sizes of a few hundred parsecs to a few Kpc have
been measured from CO and emission--line gas observations.  Since most of
the energy in luminous and ultraluminous infrared galaxies emerges in the
mid and far--infrared, this is the logical part of the spectrum from which
to judge the size of the power source.  A number of authors have presented
mid--infrared images or estimates of the compactness at $10\mu$m in some of
the nearest LIGs and ULIGs with beams ranging from $2-10\arcsec$ (Becklin
\& Wynn-Williams 1987; Matthews et al. 1987; Carico et al. 1988; Sanders et
al. 1988a, etc.) and have found a strong trend toward increasing compactness
for systems with luminosities above $\sim10^{11-12}$L$_{\odot}$.  However,
since the nearest ULIG (Arp 220) has a scale approaching 300pc/$\arcsec$,
these observations, with the exception of the drift--scan measurement of Mrk
231 by Matthews et al. (1987) which did place an upper limit of 0.5$\arcsec$
FWHM on the size of the emitting region in Mrk 231, could not rule out dense,
luminous starbursts as the source of the enormous energies.

In the near--infrared, we are measuring starlight, hot dust, and/or
non-thermal emission from a central active nucleus.  Therefore we may not
be directly probing the material responsible for the far--infrared emission
in LIGs or ULIGs.  However, the sharp and relatively stable PSF provided
by NICMOS can be used to place a limit on the maximum amount of unresolved
near--infrared light coming from the nucleus in each of our sample galaxies,
and allow us to measure the compactness of these sources on scales often
well below 100 pc.  If a significant fraction of the total near--infrared
light is unresolved at NICMOS resolution, it may provide strong evidence
for an active nucleus.

In this section, we quantify the degree of central concentration of the
light in the luminous and ultraluminous infrared galaxies in order to
investigate trends in the nuclear contribution as a function of luminosity,
merger evolutionary stage and nuclear characteristics. To accomplish this,
in a manner which is relatively independent of distance effects, we compute
the total flux with 3 kpc projected radius of the 2.2 $\mu$m peak and then
calculate the percentage of this flux contained as a function of radii less
than 3 kpc (Fig.~\ref{f_radial_kpc}). These curves of growth can then be
used to assess the contribution of a compact source (i.e., a puatative AGN)
to the total flux. The vertical bar on the top border indicates an angular
{\it{radius }} of 0.12\arcsec, corresponding to the radius (HWHM) of the
diffraction resolution at 2.2 $\mu$m.

Inspection of Fig.~\ref{f_radial_kpc} reveals that seven of the galaxies
have a significant percentage ($\geq 30$\%) of their flux originating
within $\sim 0.12\arcsec$ radius : NGC 7469, IRAS 08572+3915, IRAS
05189-2524, PKS1345+12, IRAS 07598+6508, Mrk 1014, and 3C48 (not shown
here).  All of the seven galaxies with significant nuclear point-source
contributions are also classified as warm in terms of their
mid-infrared colors, yet there are warm galaxies which do not exhibit 
significant point-like nuclei (eg. NGC 4418).  Similar conclusions can be
drawn with respect to the optical spectral classification -- i.e. most of
the galaxies with nuclear point sources contributing significantly in the
near-infrared are classified as Seyfert or QSO, yet not all of the
galaxies with AGN-like spectra have significant nuclear point sources at
the resolution of these data.  Although these qualitative correlations are
not unexpected, it should be underscored that we are here finding that,
quantitatively, a significant fraction of the flux in the near-infrared is
nucleated in these seven sources. At the same time, it is important to
recognize that even in those galaxies with strong point-like nuclei, there
is usually also a similar contribution from a component which is clearly
extended.  Since it would be very difficult to have this extended
component actually be scattered nuclear light, this extended flux is
almost certainly stellar in nature.  The half-light radii and mid-infrared
color classifications are given for the entire sample in Table 9.

A visual summary of the near--infrared nuclear concentration of our sample
LIGs and ULIGs is provided by Fig.  ~\ref{half_radius} which shows the radii
($R_{1/2}$) in which 50\% of the total flux (at $R \leq 3 $kpc) is enclosed
for all of the galaxies as a function of their far-infrared luminosities,
for both the $1.1\mu$m and $2.2\mu$m data.  (These half-light radii are not
correlated with distance of the galaxies and hence, the measured variations
are real.) As described above, nearly all of the compact galaxies, those with
R$_{1/2} < 0.5$ Kpc, are  warm (or equivalently have AGN or LINER optical
emission line classifications), although there are a few cold systems
(IRAS 12112$+$0305, NGC 2623, Zw049.057, and UGC 5101) that are nearly as
compact at $2.2\mu$m.  As a group, the cold systems show a much larger
spread in R$_{1/2}$ than do the warm systems.  Figure~\ref{half_radius}
also clearly demonstrates that there is no correlation between the total
far-infrared luminosity and the presence or absense of a significant nuclear
point-source in the near-infrared.

Lastly, it is worth noting that R$_{1/2}$ is generally smaller at
2.2 $\mu$m than at either 1.1 or 1.6 $\mu$m. This is demonstrated in
Fig.~\ref{ratio_half_radius} which shows the ratio of R$_{1/2}$ in each
band. In virtually all cases, the 1.1 or 1.6 $\mu$m sizes are similar and
their ratio is tightly clustered near unity while the ratio of the 2.2 to
1.1 $\mu$m sizes is typically $\sim 0.6$ with a larger range.  (The three
exceptions are all systems where the nuclear point source is dominant
and the larger size at 2.2 $\mu$m is simply due to the larger PSF size at
longer wavelengths.) The smaller sizes of the cores at 2.2 $\mu$m are very
likely due to the large dust extinctions at shorter wavelengths which lead
to an underestimation of the true nuclear emission component. The nuclei
are systematically much redder in color than the inner galactic disks.

\section{The Nature of the Extended Near-Infrared Light}

Both the analysis of the galaxy colors and light profiles suggest that
much of the near-infrared light from the cold galaxies is stellar in
origin.  While ground-based, near-infrared spectroscopy of luminous
infrared galaxies shows evidence for a mixture of red supergiant light,
emission from hot dust, and possibly a metal-rich giant population
(\cite{rid94}; \cite{gol95}; \cite{shi96}), the high surface brightness of
the nuclei relative to the underlying galaxy at 1--2 $\mu$m and the size
of the slits used for the near-infrared spectroscopy (3$\arcsec$) indicate
that the strong CO absorption often seen is likely coming from the nuclei
only (\cite{arm95}).

Regardless whether or not the extended light comes from stars produced as
a result of the merger, recent 8--25 imaging of two galaxies in the sample
provide strong evidence that only the high surface brightness nuclear
features observed in the near-infrared contribute significantly to the
overall bolometric luminosity of these galaxies. Mid-infrared observations
by Soifer (1998) show conclusively that the 25 $\mu$m flux of Arp 220 is
produced within the nuclear region containing the two nuclei, and Evans et
al. (1999a) conclude that the 8--25 $\mu$m emission in NGC 4418 is
compact, indicated that most of the bolometric luminosity of the galaxy is
coming from a region $\le0.5\arcsec$ in diameter. Thus, the starlight or
AGN responsible for heating the dust in these galaxies resides in the
compact nuclear regions, which in many cases, are just becoming visible at
near-infrared wavelengths.

\section {Conclusions}

The high resolution HST-NICMOS images presented here for a sample of 24
luminous infrared galaxies have revealed extremely diverse morphology,
probably reflecting their varying stages of galactic merging/interaction
and evolution. Eleven of the 24 systems exhibit double nuclei with
projected separations between 0.4 and 7 kpc. Eleven of the 24 systems also
have point-like nuclear sources and in 7 of these galaxies, the nuclear
sources produce a significant fraction ($\sim 50$\%) of the total
near-infrared flux.  All but one of the systems with significant nuclear
sources is classified as warm based on their mid-infrared colors.  These
7 galaxies also possess nuclei with red near-infrared colors indicative of
QSO light mixed with varying amounts of hot dust. With the exception of
UGC 5101 and VV114E (and possibly three others), all of the cold galaxies
have nuclear near-infrared colors consistent with reddened starlight.

The bright clusters seen in virtually every one of these galaxies have
near-infrared colors consistent with ages $\geq 5\times10^7$ yrs and their
luminosities range up to $2\times 10^{9}$ \lsun. In a few cases, they
are preferentially situated along the area of overlap of the two galactic
disks (eg. NGC 6090 and VV114) and were therefore probably formed by
hydrodynamic gas compression of the ISM. Their masses are typically a
factor of 100 greater than Galactic globular clusters and therefore they
are not simply young globular clusters. 

The relative contributions of AGNs and starlight to the near-infrared
luminosity varies among the sample galaxies.  The 7 systems
with significant nuclear point-sources are likely to have significant AGN
contributions since they also have optical emission line ratios
characteristic of a hard-spectrum ionization source. On the other hand,
for the rest of the sample, most of the near-infrared flux clearly
originates outside the central 100-300 pc, and even in those systems with
significant nuclear sources there is still approximately 50\% of the
near-infrared emanating from a spatially extended stellar population.

Nine of the 24 systems exhibit an {\it approximate} r$^{1/4}$ law
(typically over a factor of 30 in radius), suggesting that these systems
will eventually become spirals with very massive central bulges or
possibly even giant elliptical galaxies -- the latter only if the
remaining ISM is converted into stars at high efficiency.  The majority of
the sample galaxies, however, show no statistically significant preference
for either an r$^{1/4}$ or exponential surface brightness profile over the
range in radii probed by NICMOS.

\vspace{5mm}

We wish to acknowledge general contributions from many members of the
NICMOS IDT which were critical to the success of these observations. C.
Boone helped with an extensive literature search during the planning of
these observations and G.  Neugebauer, T. Soifer, J. Goldader, K.
Matthews, N. Trentham, and J. Carpenter provided helpful discussions.
Sylvain Veilleux providing very helpful suggestions as referee.  We also
thank R. Brunner and S.  Odewahn for help with Source Extractor.  The
NICMOS project has been supported by NASA grant NAG 5-3042 to the NICMOS
instrument definition team.  This paper is based on observations with the
NASA/ESA Hubble Space Telescope obtained at the Space Telescope Science
Institute, which is operated by Association of Universities for Research
in Astronomy, Incorperated, under NASA contract NAS5-26555.

\begin{appendix}{}
\section {Adaptive Filtering}

 To show both the small scale structure of bright sources and at the same
time bring out the extended low-surface brightness emission in the galaxy
envelopes, we developed a variable resolution convolution routine (in IDL)
which smooths the image with a boxcar filter with the resolution of the
boxcar depending on the local signal-to-noise ratio in the image. The
noise is measured near the sides of each image and the filter width at
each pixel was set to $\gamma\times~S(x,y)/\sigma$ where S(x,y) and
$\sigma$ are the flux at each pixel and the rms noise.  The gain constant,
$\gamma$, was set to $\sim 60$ (based on trial runs).  (Since the width
could become very large where the signal was very weak, the maximum filter
width was limited to the 10\% of image width.) For contour images
involving one wavelength, the local filter width followed the above
prescription using the flux in the individual image, but for images
involving the combination of more than one band (eg. the ratio images or
for calculating the reddening -- see below), the filter functions were
determined for each band separately and then the applied filter at each
pixel was taken to be the lowest resolution determined on either image.
This ensured that the combined image involved data at the same resolution.
In addition, for combined images, both were also smoothed by an
$0.2\arcsec$ resolution (FWHM) Gaussian filter after having been passed
through the adaptive filtering routine.

An illustration of the variable smoothing is shown in Figure
~\ref{opt_smo}. In the right panel, a contour map of the unsmoothed image
is shown, while on the left is the same data with the variable smoothing
applied. The contours levels are the identical for the two images and it
can be clearly seen that the apative filtering managed to retain all the
original detail where the signal was high but also brought out the low
level signals much more clearly.  It is noteworthy also the bright sources
in the outer part of the galaxy (where the adjacent background is low) are
also retained at the nearly the original resolution -- that is the filter
resolution adjusts quickly to compact bright sources. The adaptive filtering
used here has the advantages of being both conceptually simple and yielding
a predictable resolution.  Similar approaches have been explored before,
although primarily for processing x-ray data \cite{lorenz}; \cite{sle94};
\cite{biv96}; \cite{sur99a}).

\end{appendix}

\figcaption[lir2560.ps]{The sample galaxies are plotted as a function of
$L_{ir}$ at $\lambda$=8-1000 $\mu$m and the IRAS 25$\mu$m/60$\mu$m flux
ratio. The classification of the optical emission line ratios are
indicated by the different symbols. The vertical line at 25$\mu$m/60$\mu$m
 $= 0.2$ divides the sample into cold and warm mid-infrared
 spectral types.\label{lir2560}}

\figcaption[3color.ps]{Three color images are shown for each galaxy
arranged in order of increasing luminosity. The maximum area with
detectable emission is displayed. The blue (1.1$\mu$m), green (1.6 $\mu$m)
and red (2.2 $\mu$m) were individually log-stretched to bring out the
maximum structure and to enhance the observed color gradients.
\label{3color}}

\figcaption[4contour.ps]{Shaded contour plots are shown
for the central region of each galaxy arranged in order of increasing
luminosity. The 1.1 (upper left), 1.6 (upper right) and 2.2 (lower left)
$\mu$m images are plotted with logarithmic shading and contours spaced
by factors $2^{1/3}$.  The labelled contour values are the same for all
galaxies; the lowest level ('1') is at a surface brightnes of $10^{-5}$ Jy
arcsec $^{-2}$ and every 5'th contours is thickened and labelled (i.e. 5:
2.5$10^{-5}$, 10: 7.8$10^{-5}$, 15: 2.5$10^{-4}$, 20: 7.7$10^{-4}$, 25:
2.4.1$10^{-3}$, 30: 7.6.3$10^{-3}$, 35: 2.4$10^{-2}$ , 40: 7.5$10^{-2}$,
45: 0.23 and 50: 0.74 Jy arcsec $^{-2}$.The ratio of the $2.2$ and
$1.1$ $\mu$m images (lower right) has contours spaced by a factor of
$2^{1/3}$ starting from a value of 0.5.  The arcsec displacements in
RA and DEC, given along the borders are measured from the 2.2 $\mu$m in
all frames. At the upper left, a length bar is drawn. For the galaxies
with strong point-sources, the PSF was fit to the source and then
subtracted and replaced by a Gaussian with the proper integrated flux
(see text -- NGC 7469, IRAS 08572+3915, IRAS 05189-2524, PKS 1345+12,
IRAS 07598+6508, Mrk 1014 and 3C48). As described in the text, the image
data has been smoothed with an adaptive smoothing algorithm to reduce
the noise (and spatial resolution) in lower brightness areas but retain
the full resolution where the brightness is high (see Appendix). For
the ratio image, both the 2.2 and 1.1 $\mu$m images were smoothed with
the same adaptive smoothing and then smoothed with a Gaussian FWHM $=
0.2\arcsec $.  In cases where a strong point-source or variable background
contaminated the 2.2 $\mu$m image, the ratio was based on the 1.6 and
1.1 $\mu$m images.  \label{4contour}}

\figcaption[2contour.ps]{Shaded contour plots of the
extinction corrected 2.2 $\mu$m emission are shown together with
the 1.1 $\mu$m (upper left) observed emission. In both panels, the
contours and shading are logarithmic with the  contours spaced by factors
$2^{1/2}$. (The level values are the same as for Fig.~\ref{4contour}). The
arcsec displacements in RA and DEC, given along the borders are measured
from the 2.2 $\mu$m in all frames. At the upper left, a length bar
is drawn. For the ratio image, both the 2.2 and 1.1 $\mu$m images
were smoothed with the same adaptive smoothing and then smoothed
with a Gaussian FWHM $= 0.2\arcsec $ in calculating the 2.2 $\mu$m
opacity from Eq. 3 (see text). In cases where a strong point-source or
variable background contaminated the 2.2 $\mu$m image, the extinction
corrected image was derived for 1.6 $\mu$m. For the galaxies with strong
point-sources, the PSF was fit to the source and then subtracted and
replaced by a Gaussian with the proper integrated flux (see text -- NGC
7469, IRAS 08572+3915, IRAS 05189-2524, PKS 1345+12, IRAS 07598+6508,
Mrk 1014 and 3C48).  \label{2contour}}

\figcaption[nucflux.ps]{$m_{1.1-1.6}$ and $m_{1.6-2.2}$ color-color
diagram for the sample of galaxies with the fluxes measured in a
1.1\arcsec\ diameter aperture with the adjacent background galaxy
subtracted (i.e., `nuclear' fluxes). The different symbols denote the
luminosity and warm/cold galaxy classification. The locus for the
evolution of an instantaneous starburst with Salpeter IMF is also show.
The typical colors of free-free emission and optical PG QSOs are shown,
the latter is plotted with variable percentages of 2.2 $\mu$m emission due
to warm dust (see text). Lastly, the reddening vector based on the
extinction curve of Rieke \& Lebofsky (1985) is shown for two cases : a
foreground screen of dust (the straight vector) and a model in which the
dust is uniformly mixed with the emitting sources (the curved track). The
models were redshifted to z$= 0.05$, corresponding to a typical redshift
of the galaxies observed here.  \label{nucflux}}

\figcaption[diagnflux.ps]{Color-color diagram similar to
Fig.~\ref{nucflux} except the symbols denote the optical emission-line
classification of the galaxies.  Data for LIGs are plotted as unfilled
symbols, and the data for ULIGs are plotted as filled symbols.
\label{diagnflux}}

\figcaption[2kpcflux.ps]{Color-color diagram similar to
Fig.~\ref{nucflux} except a fixed 2 kpc-diameter aperture was used.
In the cases where the sources were too distant to use a 2 kpc-diameter
aperture (i.e., PKS 1345+12, IR 07598+6508, Mrk 1014, 3C 48), 
a 1.1$\arcsec$-diameter aperture measurement was used.
\label{2kpcflux}}


\figcaption[cluflux.ps]{Color-color diagram similar to Fig.~\ref{nucflux}
for the cluster sources with $SNR \geq 5$.  \label{cluflux}}

\figcaption[radial_kpc.ps]{Mean surface brightness (Jy $\arcsec ^{-2}$)
as a function of projected radius in arcsec (top scale) and kpc (bottom
scale) from the 2.2 $\mu$m peak in each galaxy. The 1.1, 1.6 and 2.2
$\mu$m surface brightnesses are plotted with solid, dashed and dotted lines
respectively. Only in VV114, is the surface brightness plotted separately
for the two galaxies; in some of the other objects, the secondary galaxy
contribution can be seen as a peak at large offsets (eg. IRAS 14348-1447). The
vertical bar on the top border indicates an angular radius of 0.12\arcsec
(the radius of the diffraction resolution at 2.2 $\mu$m).
 \label{radial_kpc}}

\figcaption[radial_r1_4.ps]{Mean surface brightness (Jy $\arcsec ^{-2}$)
as a function of projected $R^{1/4}$ in kpc from the 2.2 $\mu$m peak in
each galaxy. The 1.1, 1.6 and 2.2 $\mu$m surface brightnesses are plotted
with dotted, dashed and solid lines respectively. The
vertical bar on the top border indicates an angular radius of 0.12\arcsec
(the radius of the diffraction resolution at 2.2 $\mu$m).
 \label{radial_r1_4}}

\figcaption[f_radial_kpc.ps]{The percentage of the integrated flux at
projected radius 3 kpc is shown as a function of R (in Kpc).  The 1.1,
1.6 and 2.2 $\mu$m growth curves are plotted with dotted, dashed, and solid
lines respectively. The warm(W) or cold(C) classification of the mid-infrared
emission of the galaxy is noted in the upper left and the vertical bar on
the top border indicates an angular radius of 0.12\arcsec (the radius of
the diffraction resolution at 2.2 $\mu$m).  \label{f_radial_kpc}}

\figcaption[half_radius.ps]{The radius containing 50\% of the total flux
inside 3 kpc is plotted as a function of $L_{ir}$ for galaxies with
warm (W) and cold (C) mid-infrared color classification. The
upper panel shows the half-light radius for the 1.1 $\mu$m emission and
the lower panel that for the 2.2 $\mu$m emission.  \label{half_radius}}

\figcaption[ratio_half_radius.ps]{The ratios of half-light radii at 1.6 to
1.1 $\mu$m (upper panel) and 2.2 to 1.1 $\mu$m (lower panel) are shown as a
function of $L_{ir}$ for galaxies with warm (W) and cold (C) mid-infrared
color classification. The two warm objects with large R(2.2)/R(1.1) are
simply unresolved. \label{ratio_half_radius}}

\figcaption[opt_smo.ps]{Example of the use of the variable smoothing
algorithm as applied to the 1.1 and 2.2 $\mu$m images for IRAS
17208-0014.  In the upper panels are contours drawn from the original
images and on the bottom contours drawn after the adaptive smoothing was
applied. The contours a logarithmic, spaced by factors $2^{1/2}$ and are
the same for all four panels. No attempt was made to remove PSF effects
and the color fringes around such sources are due to the increasing width
of the PSF at longer wavelengths (see text).\label{opt_smo}}

\vfill\eject

\begin{deluxetable}{lcccccc}
\scriptsize
\def\arcsec{\hbox{$^{\prime\prime}$}}
\tablenum{1}       
\tablewidth{0pt}
\tablecaption{Luminous IR Galaxies Observed w/ NICMOS}
\tablehead{
\colhead{Name } &
\colhead{Nuclear Spec.} &
\colhead{ IR Class.} &
\colhead{z} &
\colhead{ log $L_{ir}$ } &
\colhead{ log M(H$_2$) } &
\colhead{ NIR Morphology} \nl
}
\startdata
NGC 4418&\nodata$^1$&W&.007&11.0&9.00&Nuclear dust disk\nl

Zw049.057&H II$^a$&C&.0131&11.22&8.78&Inclined dusty disk\nl

NGC 6090&H II$^a$&C&.0294&11.51&10.15&Double Galaxy, clusters and Pt.
Nuc.\nl

NGC 2623&\nodata$^1$&C&.0185& 11.54&9.77&Spiral w/ clusters \nl

IC 883&LINER$^a$&C&.0231&11.60&9.87&Disk w/ dusty spiral\nl

NGC 7469&Seyfert 1$^a$&W&.0166&11.60&9.96&Pt. nucleus w/ spiral\nl

VV 114&H II$^a$&C&.0201&11.62&10.44&Double galaxies\nl

NGC 6240&LINER$^a$&C&.0243&11.82&10.29&Double nuclei ($1.6\arcsec$)\nl

VIIZw031&H II$^b$&C&.0542&11.94&10.69&Asymmetric Spiral\nl

IRAS 15250+3609&LINER$^a$&C&.0534&12.00&\nodata&Double nuclei(?)
($0.7\arcsec$) + clusters\nl

UGC 5101&LINER$^a$&C&.0400&12.01&10.44&Pt. nucleus with disk \nl

IRAS 10565+2448&H II$^a$&C&.0430&12.02&10.34&Double galaxies ($8\arcsec$)
w/ tail\nl

IRAS 08572+3915&LINER$^a$&W&.0582&12.09&9.79&Double galaxies ($5\arcsec$)
w/ tail\nl

IRAS 05189-2524&QSO$^{a,2,3}$&W&.0427&12.10&10.37&Pt. nucleus\nl

IRAS 22491-1808&H II$^a$&C&.0773&12.10&10.43&Double nuclei ($1.6\arcsec$)
+ tails\nl

Mrk 273&Seyfert 2$^c$&C&.0378&12.11&10.24&Pt. double nuclei
($1\arcsec$)\nl

Arp 220&LINER$^c$&C&.0185&12.19&10.00&Double nuclei ($1\arcsec$) w/
dust\nl

PKS 1345+12&Seyfert 2$^{c,2}$&W&.1224&12.22&10.78&Double nuclei
($3\arcsec$)\nl

IRAS 12112+0305&LINER$^c$&C&.0727&12.26&10.62&Double nuclei ($3\arcsec$)
w/ tail\nl

IRAS 14348-1447&LINER$^a$&C&.0825&12.27&10.78&Double Nuclei
(3.5\arcsec)\nl

IRAS 17208-0014&H II$^a$&C&.0429&12.40&10.71&Nuclear disk w/ clusters\nl

IRAS 07598+6508&Seyfert 1$^{c,4}$&W&.149&12.45&10.73&Pt. nucleus +
extended structure\nl

Mrk 1014&QSO$^d$&W&.163&12.49&10.61&QSO + spiral disk\nl

3C 48&QSR$^d$&W&.398&12.50&10.55&QSO with extended structure\nl

\enddata

\tablerefs{Emission line classifications: (a) Veilleux, Kim \& Sanders
(1999); (b) Djorgovski et al. (1990); (c) Kim, Veilleux, \& Sanders (1998);
(d) Schmidt \& Green (1983).}

\tablenotetext{1}{Reliable optical emission-line classifications of these
two galaxies do not exist.}

\tablenotetext{2}{Veilleux, Sanders \& Kim (1999) have detected
broad Pa$\alpha$ emission from these galaxies, indicating that it contains
a buried quasar nucleus.}

\tablenotetext{3}{Young et al. (1996) have detected a QSO-like broad-line
region in scattered light.}

\tablenotetext{4}{Hines \& Wills (1995) classify this as a BALQSO.}

\end{deluxetable}

\begin{deluxetable}{llrrr}
\tablenum{2}
\tablewidth{0 pt}
\tablecaption{Journal of Observations}
\tablehead{
\colhead{Name} &
\multicolumn{1}{c}{Date} &
\multicolumn{3}{c}{Integration Time (sec)}
\nl
\colhead{} & 
\multicolumn{1}{c}{} & 
\multicolumn{1}{c}{1.1 $\mu$m} &
\multicolumn{1}{c}{1.6 $\mu$m} &
\multicolumn{1}{c}{2.2 $\mu$m}
} 
\startdata
NGC 4418      & 1997 Nov 26 & 352 & 352 & 480 \nl
Zw049.057     & 1997 Dec 29 & 160 & 224 & 256 \nl
NGC 6090      & 1997 Nov 10 & 384 & 384 & 544 \nl
NGC 2623      & 1997 Nov 19 & 352 & 352 & 480 \nl
IC 883        & 1997 Nov 21 & 352 & 352 & 480 \nl
NGC 7469\tablenotemark{a}      & 1997 Nov 10 & 352 & 352 & 480 \nl
VV 114E       & 1998 Aug 03 & 224 & 224 & 280 \nl
VV 114W       & 1998 Aug 03 & 224 & 224 & 280 \nl
NGC 6240\tablenotemark{a}      & 1998 Feb 12 & 160 & 192 & 224 \nl
VII Zw 31     & 1997 Nov 17 & 600 & 600 & 680 \nl
IRAS 15250+3609\tablenotemark{b} & 1997 Nov 19 & 224 & 224 & 320 \nl
UGC 5101      & 1997 Nov 07 & 560 & 560 & 680 \nl
IRAS 10565+2448 & 1997 Nov 29 & 480 & 480 & 600 \nl
IRAS 08572+3915\tablenotemark{b} & 1997 Nov 11 & 224 & 160 & 288 \nl
IRAS 05189-2524\tablenotemark{b} & 1997 Dec 09 & 224 & 224 & 288 \nl
IRAS 22491-1808 & 1997 Nov 21 & 480 & 480 & 600 \nl
Mrk 273\tablenotemark{b}       & 1997 Dec 10 & 256 & 256 & 320 \nl
Arp 220\tablenotemark{b}       & 1997 Apr 04 & 1024 & 1024 & 1024 \nl    
PKS 1345+12   & 1997 Dec 05 & 480 & 480 & 600 \nl
IRAS 12112+0305\tablenotemark{b} & 1997 Nov 15 & 192 & 192 & 224 \nl
IRAS 14348-1447 & 1997 Dec 31 & 480 & 480 & 600 \nl
IRAS 17208-0014\tablenotemark{b} & 1997 Oct 26 & 224 & 224 & 288 \nl
IRAS 07598+6508 & 1997 Nov 11 & 224 & 224 & 256 \nl
Mrk 1014      & 1997 Dec 13 & 480 & 480 & 600 \nl
3C 48         & 1997 Dec 11 & 480 & 480 & 600 \nl
\enddata
\tablenotetext{a}{2.2 $\mu$m Observations of a PSF star obtained during
this orbit.}
\tablenotetext{b}{1.1, 1.6, and 2.2 $\mu$m observations of a PSF star
obtained during this orbit.}
\end{deluxetable}

\begin{deluxetable}{lccccc}
\tablenum{3}       
\tablewidth{0pt}
\tablecaption{Morphological Features in IR Luminous Galaxies}
\tablehead{
\colhead{Name } &
\colhead{Nuclear Pt. Source\tablenotemark{a}} &
\colhead{ Nuclear} &
\colhead{Clusters} &
\colhead{ Spiral\tablenotemark{b}} &
\colhead{ Nuclear } \nl
\colhead{} &
\colhead{} &
\colhead{Separation} &
\colhead{} &
\colhead{} &
\colhead{Dust\tablenotemark{c}} \nl
\colhead{} &
\colhead{ ~~ 1.1~ 1.6~ 2.2$\mu$m } &
\colhead{\arcsec ~~~ kpc} &
\colhead{} &
\colhead{} & \nl
}
\startdata
NGC 4418&  &&  &  ? &Y\nl
Zw049.057&  &&Y& Y&Y\nl
NGC 6090& Y ~ Y ~ Y & 6.0\arcsec ~~~ 3.4& Y & Y &\nl
NGC 2623&  &&Y&Y &\nl
IC 883&  &&Y& &Y\nl
NGC 7469& Y ~ Y ~ Y &80\arcsec ~~~ 26&Y&Y &\nl
VV 114&  & 14\arcsec ~~~ 4.7& Y & Y &\nl
NGC 6240&  &$1.6\arcsec$ ~~~ 0.8&Y& &Y\nl
VIIZw031&  &&Y&Y &\nl
IRAS 15250+36&  & ($0.7\arcsec$ ~~~ 0.7)&Y& &\nl
UGC 5101& ~~~~~~~~~ Y &  & Y&Y &\nl
IRAS 10565+2448&  &$8.0\arcsec$ ~~~ 6.7&Y & Y &\nl
IRAS 08572+3915& Y ~ Y ~ Y &$5.0\arcsec$ ~~~ 5.6&Y& &\nl
IRAS 05189-2524& Y ~ Y ~ Y & & & &\nl
IRAS 22491-1808&  &$1.6\arcsec$ ~~~ 2.4& Y& &Y\nl
Mrk 273& ~ ~ ~ ~ ~~Y &$1.0\arcsec$ ~~~ 0.7&Y& &Y\nl
Arp 220&  &$0.9\arcsec$ ~~~ 0.4& Y& &Y\nl
PKS 1345+12& Y ~ Y ~ Y &$3.0\arcsec$ ~~~ 7.1& & &Y\nl
IRAS 12112+0305& ~ ~ ~ ~ ~~Y &$3.0\arcsec$ ~~~ 4.2& & &Y\nl
IRAS 14348-1447&  &$3.5\arcsec$ ~~~ 5.6& Y & Y &\nl
IRAS 17208-0014&  && Y & &\nl
IRAS 07598+6508& Y ~ Y ~ Y & & &\nl
Mrk 1014& Y ~ Y ~ Y  &  &  & Y&\nl
3C 48& Y ~ Y ~ Y & &  &\nl
\enddata
\tablenotetext{a}{Point source at resolutions 0.1 -- 0.2 \arcsec at 1.1 --
2.2 $\mu$m}
\tablenotetext{b}{Spiral Arms seen in central 100 pc to 1 kpc.}
\tablenotetext{c}{Strongly variable reddenning in nucleus.}
\end{deluxetable}

\begin{deluxetable}{lrrrrrr}
\tablenum{4}
\tablewidth{0pt}
\tablecaption{Aperture Photometry}
\tablehead{
\multicolumn{1}{c}{Name} &
\multicolumn{1}{c}{Aperture} &
\multicolumn{1}{c}{m$_{1.1}$} &
\multicolumn{1}{c}{m$_{1.6}$} &
\multicolumn{1}{c}{m$_{2.2}$} &
\multicolumn{1}{c}{m$_{1.1}$ - m$_{1.6}$} &
\multicolumn{1}{c}{m$_{1.6}$ - m$_{2.2}$}\nl
\multicolumn{1}{c}{} &
\multicolumn{1}{c}{(Diameter)} &
\multicolumn{1}{c}{} &
\multicolumn{1}{c}{} &
\multicolumn{1}{c}{} &
\multicolumn{1}{c}{} &
\multicolumn{1}{c}{}}
\startdata

NGC 4418 & 1.1$\arcsec$ & 14.98 & 13.85 & 13.26 & 1.13 & 0.59 \nl 
   & 5.0$\arcsec$ & 13.22 & 12.21 & 11.77 & 1.00 & 0.44 \nl
   & 11.4$\arcsec$ & 12.54 & 11.62 & 11.25 & 0.91 & 0.36 \nl
\hline

Zw049.057 & 1.1$\arcsec$ & 16.02 & 14.72 & 13.98 & 1.30 & 0.74 \nl
   & 5.0$\arcsec$ & 13.88 & 12.69 & 12.16 & 1.19 & 0.54 \nl
   & 11.4$\arcsec$ & 10.25 & 9.33 & 8.48 & 0.92 & 0.85  \nl
\hline

NGC 6090E & 1.1$\arcsec$ & 15.49 & 14.51 & 14.04 & 0.99 & 0.47  \nl
   & 5.0$\arcsec$ & 13.46 & 12.52 & 11.96 & 0.94 & 0.56  \nl
   & 11.4$\arcsec$ & 12.93 & 12.00 & 11.40 & 0.93 & 0.60 \nl
\hline

NGC 6090W & 1.1$\arcsec$& 16.45 & 15.62 & 15.19 & 0.83 & 0.43 \nl
   &  5.0$\arcsec$ & 14.59 & 13.79 & 13.42 & 0.80 & 0.38 \nl
\hline

NGC 2623 & 1.1$\arcsec$ & 15.10  & 13.29  & 12.10  & 1.81 & 1.19  \nl
   & 5.0$\arcsec$ & 13.52  & 12.16 & 11.37 & 1.36  & 0.79  \nl
   & 11.4$\arcsec$ & 12.80 & 11.58 & 11.08 & 1.22 & 0.49 \nl
\hline

IC 883 & 1.1$\arcsec$ & 16.08 & 14.32 & 13.11 & 1.76 & 1.21 \nl
   & 5.0$\arcsec$ & 13.92 & 12.58 & 11.69 & 1.34 & 0.88 \nl
   & 11.4$\arcsec$ & 13.16 & 12.00 & 11.24 & 1.16 & 0.76 \nl
\hline

NGC 7469 & 1.1$\arcsec$ & 12.41 & 11.14 & 9.89 & 1.27 & 1.25 \nl
   & 5.0$\arcsec$ & 11.43 & 10.23 & 9.23 & 1.20 & 1.00  \nl
   & 11.4$\arcsec$ & 11.14 & 9.96 & 9.05 & 1.18 & 0.92 \nl
\hline

VV 114E & 1.1$\arcsec$ & 16.06 & 14.34 & 12.62 & 1.72 & 1.72 \nl
   & 5.0$\arcsec$ & 13.63 & 12.22 & 11.25 & 1.41 & 0.97 \nl 
   & 11.4$\arcsec$ & 12.87 & 11.63 & 10.80 & 1.24 & 0.83 \nl 
\hline

VV 114W & 1.1$\arcsec$ & 15.63 & 14.85 & 14.52 & 0.78 & 0.33 \nl
   & 5.0$\arcsec$ & 13.46 & 12.65 & 12.30 & 0.81 & 0.32 \nl
   & 11.4$\arcsec$ & 12.66 & 11.82 & 11.47 & 0.84 & 0.35 \nl
\hline

NGC 6240S & 1.1$\arcsec$ & 13.75 & 12.25 & 11.36 & 1.50 & 0.89 \nl
   & 5.0$\arcsec$ & 12.61 & 11.18 & 10.34 & 1.42 & 0.84 \nl
   & 11.4$\arcsec$ & 12.11 & 10.72 & 9.92 & 1.39 & 0.80  \nl
\hline 

NGC 6240N & 1.1$\arcsec$ & 14.83 & 13.45 & 12.72 & 1.38 & 0.74 \nl
   & 5.0$\arcsec$ & 12.62 & 11.19 & 10.36 & 1.42 & 0.83 \nl
   & 11.4$\arcsec$ & 12.10 & 10.71 & 9.91 & 1.39 & 0.80  \nl
\hline

VIIZw031 & 1.1$\arcsec$ & 15.33 & 14.05 & 13.29 & 1.28 & 0.75 \nl
  & 5.0$\arcsec$ & 13.65 & 12.43 & 11.68 & 1.21 & 0.76 \nl
  & 11.4$\arcsec$ & 13.21 & 12.06 & 11.22 & 1.15 & 0.83 \nl
\hline

IRAS 15250+3609 & 1.1$\arcsec$ & 16.18 & 14.90 & 14.06 & 1.27 & 0.84 \nl
   &  5.0$\arcsec$ & 14.63 & 13.58 & 12.99 & 1.05 & 0.59 \nl
   &  11.4$\arcsec$ & 14.46 & 13.41 & 12.77 & 1.06 & 0.64 \nl
\hline

UGC 5101 & 1.1$\arcsec$ & 14.92 & 13.32 & 11.77 & 1.61 & 1.55 \nl
   &  5.0$\arcsec$ & 13.66 & 12.25 & 11.09 & 1.41 & 1.16 \nl
   &  11.4$\arcsec$ & 13.22 & 11.88 & 10.84 & 1.34 & 1.04 \nl
\hline

IRAS 10565+2448 & 1.1$\arcsec$ & 14.84 & 13.42 & 12.49 & 1.42 & 0.93 \nl
   &  5.0$\arcsec$ & 13.64 & 12.38 & 11.60 & 1.26 & 0.79 \nl
   &  11.4$\arcsec$ & 13.38 & 12.18 & 11.46 & 1.20 & 0.72 \nl
\hline

IRAS 08572+3915N & 1.1$\arcsec$ & 17.45 & 15.80 & 13.53 & 1.65 & 2.27 \nl
   & 5.0$\arcsec$ & 16.21 & 14.80 & 13.20 & 1.42 & 1.59 \nl
   & 11.4$\arcsec$ & 15.85 & 14.20 & 13.06 & 1.64 & 1.15 \nl
\hline

IRAS 08572+3915S & 1.1$\arcsec$& 16.87 & 15.86 & 15.76 & 1.01 & 0.11 \nl 
    & 5.0$\arcsec$ & 16.68 & 15.64 & 15.74 & 1.04 & -0.09  \nl
\hline

IRAS 22491-1808W & 1.1$\arcsec$ & 16.61 & 15.62 & 15.08 & 0.99 & 0.54 \nl
   & 5.0$\arcsec$ &  15.09 & 14.06 & 13.59 & 1.04 & 0.46 \nl
   & 11.4$\arcsec$ & 14.67 & 13.71 & 13.70 & 0.96 & 0.01 \nl
\hline

IRAS 22491-1808E & 1.1$\arcsec$ & 17.34 & 16.11 & 15.37 & 1.23 & 0.74 \nl
\hline

IRAS 05189-2524 & 1.1$\arcsec$ & 13.56 & 11.83 & 10.33 & 1.73 & 1.50 \nl
   & 5.0$\arcsec$ & 13.06 & 11.48 & 10.08 & 1.58 & 1.40 \nl
   & 11.4$\arcsec$ & 13.02 & 11.40 & 10.02 & 1.61 & 1.38 \nl
\hline

Mrk 273S & 1.1$\arcsec$& 16.04 & 14.52 & 13.29 & 1.52 & 1.22 \nl
   & 5.0$\arcsec$ & 13.60 & 12.36 & 11.55 & 1.24 & 0.80 \nl
   & 11.4$\arcsec$ & 12.94 & 11.79 & 11.15 & 1.14 & 0.64 \nl
\hline

Mrk 273N & 1.1$\arcsec$& 15.36 & 13.93 & 12.91 & 1.44 & 1.02 \nl
   & 5.0$\arcsec$ &  13.56 & 12.33 & 11.53 & 1.23 & 0.80 \nl
\hline

Arp 220W & 1.1$\arcsec$ & 15.93 & 14.00 & 12.79 & 1.93 & 1.21 \nl 
   & 5.0$\arcsec$ & 13.76 & 12.21 & 11.22 & 1.55 & 0.99  \nl
   & 11.4$\arcsec$ & 12.84 & 11.54 & 10.74 & 1.30 & 0.80 \nl
\hline

Arp 220E & 1.1$\arcsec$ & 16.50 & 14.49 & 13.13 & 2.01 & 1.36 \nl
\hline

PKS 1345+12W & 1.1$\arcsec$ & 16.66 & 15.45 & 13.96 & 1.20 & 1.49 \nl
   & 5.0$\arcsec$ & 15.10 & 14.00 & 13.02 & 1.10 & 0.98 \nl
   & 11.4$\arcsec$ & 14.42 & 13.34 & 12.42 & 1.07 & 0.92 \nl
\hline

PKS 1345+12E & 1.1$\arcsec$& 16.86 & 15.85 & 15.31 & 1.01 & 0.54 \nl
\hline

IRAS 12112+0305S & 1.1$\arcsec$ & 16.88 & 15.37 & 14.37 & 1.51 & 1.00  \nl
   & 5.0$\arcsec$ & 15.74 & 14.42 & 13.72 & 1.32 & 0.70 \nl
   & 11.4$\arcsec$ & 14.86 & 13.58 & 12.96 & 1.28 & 0.62 \nl
\hline

IRAS12112+0305N & 1.1$\arcsec$& 15.81 & 14.62 & 13.89 & 1.19 & 0.72 \nl
\hline

IRAS 14348-1447S & 1.1$\arcsec$ & 16.58 & 15.09 & 14.19 & 1.49 & 0.90 \nl
   & 5.0$\arcsec$ & 15.23 & 14.02 & 13.35 & 1.22 & 0.66 \nl
   & 11.4$\arcsec$ & 14.63 & 13.45 & 12.91 & 1.18 & 0.54 \nl
\hline

IRAS 14348-1447N & 1.1$\arcsec$& 17.24 & 15.71 & 14.92 & 1.53 & 0.79 \nl
    & 5.0$\arcsec$ & 15.70 & 14.36 & 13.86 & 1.35 & 0.50 \nl
\hline

IRAS 17208-0014 & 1.1$\arcsec$ & 16.15 & 14.33 & 13.12 & 1.82 & 1.21 \nl
   & 5.0$\arcsec$ & 14.02 & 12.58 & 11.74 & 1.44 & 0.83 \nl
   & 11.4$\arcsec$ & 13.38 & 12.09 & 11.49 & 1.30 & 0.60 \nl
\hline

IRAS 07598+6508 & 1.1$\arcsec$ & 13.40 & 12.02 & 10.51 & 1.38 & 1.51 \nl
   & 5.0$\arcsec$ & 13.24 & 11.82 & 10.34 & 1.42 & 1.48 \nl
   & 11.4$\arcsec$ & 13.23 & 11.78 & 10.34 & 1.45 & 1.44 \nl
\hline

Mrk 1014 &  1.1$\arcsec$& 14.80 & 13.56 & 12.25 & 1.24 & 1.31 \nl
   & 5.0$\arcsec$ & 14.41 & 13.16 & 11.98 & 1.25 & 1.18 \nl
   & 11.4$\arcsec$ & 14.37 & 13.04 & 11.97 & 1.34 & 1.075 \nl
\hline

3C 48  & 1.1$\arcsec$ &  14.87 & 14.00 & 12.82 & 0.87 & 1.18 \nl
   & 5.0$\arcsec$ & 14.71 & 13.76 & 12.64 & 0.95 & 1.12 \nl
   & 11.4$\arcsec$ & 14.73 & 13.74 & 12.65 & 0.99 & 1.08 \nl

\enddata
\end{deluxetable}

\newcommand{\ts}{\thinspace}
\begin{deluxetable}{lrrrrr}
\tablenum{5}
\tablewidth{0pt}
\tablecaption{Nuclear Photometry\tablenotemark{a}}
\tablehead{
\multicolumn{1}{c}{Name} &
\multicolumn{1}{c}{m$_{1.1}$} &
\multicolumn{1}{c}{m$_{1.6}$} &
\multicolumn{1}{c}{m$_{2.2}$} &
\multicolumn{1}{c}{m$_{1.1}$ - m$_{1.6}$} &
\multicolumn{1}{c}{m$_{1.6}$ - m$_{2.2}$}\nl
\multicolumn{1}{c}{} &
\multicolumn{1}{c}{} &
\multicolumn{1}{c}{} &
\multicolumn{1}{c}{} &
\multicolumn{1}{c}{} &
\multicolumn{1}{c}{}}
\startdata

NGC 4418*  & 16.04 & 14.78 & 14.05 & 1.26 & 0.73 \nl 

Zw049.057*  & 17.18 & 15.98 & 15.20 & 1.20 & 0.78 \nl

NGC 6090E*  & 16.37 & 15.40 & 14.94 & 0.97 & 0.45 \nl

NGC 6090W* & 17.26 & 16.40 & 15.94 & 0.85 & 0.47 \nl

NGC 2623*  & 15.69 & 13.65 & 12.34 & 2.04 & 1.31 \nl 

IC 883*  & 17.20 & 15.02 & 13.66 & 2.18 & 1.36 \nl 

NGC 7469  & 12.41 & 11.14 & 9.89 & 1.27 & 1.25 \nl

NGC 6240S* & 14.03 & 12.53 & 11.66 & 1.50 & 0.86 \nl 

NGC 6240N* & 15.60 & 14.31 & 13.69 & 1.28 & 0.62 \nl

VIIZw031* & 16.07 & 14.80 & 14.08 & 1.28 & 0.71 \nl

IRAS 15250+3609* & 16.94 & 15.39 & 14.38 & 1.55 & 1.01 \nl

UGC 5101* & 15.44 & 13.79 & 12.04 & 1.64 & 1.75 \nl

IRAS 10565+2448* & 15.62 & 14.05 & 13.02 & 1.57 & 1.04 \nl 

IRAS 08572+3915N  & 17.45 & 15.80 & 13.53 & 1.65 & 2.27 \nl

IRAS 08572+3915S & 16.87 & 15.86 & 15.76 & 1.01 & 0.11 \nl 

IRAS 22491-1808W* & 16.92 & 15.94 & 15.37 & 0.98 & 0.57 \nl

IRAS 22491-1808E* & 18.55 & 17.13 & 16.16 & 1.42 & 0.97 \nl

IRAS 05189-2524  & 13.56 & 11.83 & 10.33 & 1.73 & 1.50 \nl

Mrk 273S* & 17.29 & 15.29 & 13.77 & 2.01 & 1.51 \nl

Mrk 273N* & 16.04 & 14.23 & 13.25 & 1.81 & 0.98 \nl

Arp 220W* & 17.13 & 14.88 & 13.48 & 2.25 & 1.40 \nl 

Arp 220E* & 17.86 & 15.94 & 14.43 & 1.92 & 1.51 \nl

PKS 1345+12W* & 17.06 & 15.85 & 14.14 & 1.21 & 1.71\nl

PKS 1345+12E* & 17.39 & 16.41 & 15.91 & 0.98 & 0.50 \nl

IRAS 12112+0305S* & 17.19 & 15.60 & 14.51 & 1.58 & 1.09 \nl 

IRAS 14348-1447S* & 17.05 & 15.36 & 14.49 & 1.69 & 0.87 \nl

IRAS 14348-1447N* & 18.02 & 16.00 & 15.31 & 2.03 & 0.69 \nl

IRAS 17208-0014* & 17.20 & 15.30 & 13.85 & 1.90 & 1.44 \nl  

IRAS 07598+6508  & 13.40 & 12.02 & 10.51 & 1.38 & 1.51 \nl

Mrk 1014 &   14.80 & 13.56 & 12.25 & 1.24 & 1.31 \nl

3C 48   &  14.87 & 14.00 & 12.82 & 0.87 & 1.18 \nl

\enddata
\tablenotetext{a}{Measured in a 1.1$\arcsec$ aperture.}
\tablenotetext{*}{Underlying galaxy subtraction performed.}
\end{deluxetable}

\begin{deluxetable}{lrrrrr}
\tablenum{6}
\tablewidth{0pt}
\tablecaption{2 Kpc-Diameter Aperture Photometry}
\tablehead{
\multicolumn{1}{c}{Name} &
\multicolumn{1}{c}{m$_{1.1}$} &
\multicolumn{1}{c}{m$_{1.6}$} &
\multicolumn{1}{c}{m$_{2.2}$} &
\multicolumn{1}{c}{m$_{1.1}$ - m$_{1.6}$} &
\multicolumn{1}{c}{m$_{1.6}$ - m$_{2.2}$}\nl
\multicolumn{1}{c}{} &
\multicolumn{1}{c}{} &
\multicolumn{1}{c}{} &
\multicolumn{1}{c}{} &
\multicolumn{1}{c}{} &
\multicolumn{1}{c}{}}
\startdata

NGC 4418 &  12.33  &    11.46   &   11.14  & 0.87  &  0.32 \nl
Zw049.057 &  13.39 &     12.24  &    11.81 &  1.15 &   0.43 \nl
NGC 6090E & 13.82  &    12.89  &  12.38 &  0.92  &  0.51 \nl
NGC 6090W & 14.98  &    14.17  &  13.76 &  0.82  &  0.41 \nl
NGC 2623  & 13.40  &    12.06  &  11.34 &  1.33  &  0.73 \nl
IC 883  & 14.02   &   12.65  &  11.77 &  1.36  &  0.88 \nl
NGC 7469 & 11.34 &     10.14 &   9.17 &   1.20 &    0.97 \nl
VV114E &  13.54  &    12.15  &  11.22 &  1.39  &   0.93 \nl
VV114W  & 13.39  &    12.57  &  12.24 &  0.82  &   0.33 \nl
NGC 6240N &  12.85  &    11.39  &  10.56 &  1.46  & 0.84 \nl
NGC 6240S  & 12.71  &    11.28  &  10.46 &   1.43 &   0.82 \nl
VIIZw031   & 14.59  &    13.32  &  12.57 &   1.27 &   0.75 \nl
IR 15250+3609 & 15.41  &    14.28  &  13.59  &  1.13  &  0.69 \nl
UGC 5101 & 14.12   &   12.61  &  11.33  &  1.51  &  1.28 \nl
IR 10565+2448 & 14.02   &   12.69  &  11.86   & 1.33   & 0.83 \nl
IR 08572+3915N & 16.92  &    15.42 &   13.42  &  1.50  &  2.00  \nl
IR 08572+3915S & 17.44  &    16.45 &   15.60  &   0.99 &   0.85 \nl
IR 22491-1808W & 16.42  &    15.42 &   14.92  &  0.99  &  0.50 \nl
IR 22491-1808E & 16.91  &    15.72 &    15.05 &   1.19 &   0.67 \nl
IR 05189-2524 &  13.19  &    11.59 &   10.19  &  1.60  &  1.40 \nl
Mrk 273N & 14.20    &  12.87  &  11.94  &  1.33  &  0.93 \nl
Mrk 273S & 14.26   &   12.90  &  11.97  &  1.35  &  0.93 \nl
Arp 220 &  13.61  &     12.10 &   11.16 &   1.51 &   0.94 \nl
IR 12112-0305N & 16.66    &  15.20  &   14.28 &   1.46 &   0.92 \nl
IR 12112-0305S & 17.04     & 15.68  &  14.79  &  1.36  &  0.89 \nl
IR 14348-1447N &  17.01    &  15.65  &  14.80  &  1.36  &   0.85 \nl
IR 14348-1447S &  16.42   &   15.01  &  14.10  &  1.41  &  0.91 \nl
IR 17208-0014 &  14.82  &    13.20  &  12.21  &  1.62  &  0.98 \nl
\enddata
\end{deluxetable}

\begin{deluxetable}{lrrrrrrrrrr}
\scriptsize
\tablenum{7}
\tablewidth{0pt}
\tablecaption{Cluster Photometry}
\tablehead{
\multicolumn{1}{c}{Number} &
\multicolumn{1}{c}{m$_{1.1}$} &
\multicolumn{1}{c}{$\Delta {\rm m}_{1.1}$\tablenotemark{b}} &
\multicolumn{1}{c}{m$_{1.6}$} &
\multicolumn{1}{c}{$\Delta {\rm m}_{1.6}$\tablenotemark{b}} &
\multicolumn{1}{c}{m$_{2.2}$} &
\multicolumn{1}{c}{$\Delta {\rm m}_{2.2}$\tablenotemark{b}} &
\multicolumn{1}{c}{m$_{1.1}$ - m$_{1.6}$} &
\multicolumn{1}{c}{m$_{1.6}$ - m$_{2.2}$} &
\multicolumn{1}{c}{N Offset\tablenotemark{c}} &
\multicolumn{1}{c}{E Offset\tablenotemark{c}}\nl
\multicolumn{1}{c}{} &
\multicolumn{1}{c}{} &
\multicolumn{1}{c}{} &
\multicolumn{1}{c}{} &
\multicolumn{1}{c}{} &
\multicolumn{1}{c}{} &
\multicolumn{1}{c}{} &
\multicolumn{1}{c}{} &
\multicolumn{1}{c}{} &
\multicolumn{1}{c}{($\arcsec$)} &
\multicolumn{1}{c}{($\arcsec$)}\nl}
\startdata
 
\multicolumn{11}{c}{Zw049.057}\nl
\hline

 01 & 19.68 & 0.16 &  18.71 & 0.10 & $>$18.55\tablenotemark{a} & 0.00 & 0.97 & 0.00 &  0.3 &  1.1 \nl 
 02 & 20.16 & 0.24 &  19.53 & 0.20 & $>$16.52 & 0.00 & 0.63 & 0.00 &  0.4 &  0.9 \nl 
\hline

\multicolumn{11}{c}{NGC 6090E}\nl
\hline

  01 & 18.26 & 0.02 &  18.15 & 0.03 &  18.20 & 0.15 & 0.11 & -.05 & -1.6 &  0.4  \nl
  02 & 18.69 & 0.03 &  18.58 & 0.04 &  18.73 & 0.23 & 0.11 & -.14 & -1.2 & -0.3  \nl
  03 & 19.14 & 0.04 &  19.23 & 0.08 & $>$19.09 & 0.00 & -.09 & 0.00 & -1.0 & -0.7  \nl
  04 & 19.19 & 0.05 &  18.83 & 0.05 &  18.59 & 0.21 & 0.36 & 0.24 & -5.1 &  3.6  \nl
  05 & 19.39 & 0.06 &  18.84 & 0.05 &  18.74 & 0.23 & 0.56 & 0.10 & -5.1 &  3.3  \nl
  06 & 19.46 & 0.06 &  18.34 & 0.03 &  17.47 & 0.08 & 1.13 & 0.87 & -1.9 & -2.0  \nl
  07 & 19.64 & 0.07 &  18.90 & 0.06 &  19.01 & 0.29 & 0.73 & -.11 & -5.0 &  4.7  \nl
  08 & 19.70 & 0.07 &  18.59 & 0.04 &  18.54 & 0.20 & 1.11 & 0.05 & -1.0 & -1.4  \nl
  09 & 19.76 & 0.08 &  19.42 & 0.09 & $>$19.09 & 0.00 & 0.33 & 0.00 & -0.4 & -1.0  \nl
  10 & 19.88 & 0.09 &  19.46 & 0.10 & $>$19.08 & 0.00 & 0.42 & 0.00 & -5.4 &  3.8  \nl
  11 & 19.91 & 0.09 &  19.18 & 0.07 &  18.88 & 0.26 & 0.73 & 0.30 & -4.9 &  5.0  \nl
  12 & 19.99 & 0.09 &  19.41 & 0.09 &  18.83 & 0.25 & 0.58 & 0.57 &  0.4 & -0.8   \nl
  13 & 20.13 & 0.11 &  19.19 & 0.07 & $>$19.10 & 0.00 & 0.94 & 0.00 &  1.7 & -0.1  \nl
  14 & 20.19 & 0.11 &  19.47 & 0.10 & $>$19.09 & 0.00 & 0.72 & 0.00 & -5.2 &  3.2  \nl
  15 & 20.32 & 0.13 &  19.56 & 0.10 & $>$19.01 & 0.00 & 0.76 & 0.00 & -1.2 & -1.4  \nl
  16 & 20.33 & 0.13 &  19.80 & 0.13 & $>$19.10 & 0.00 & 0.53 & 0.00 & -4.8 &  2.5  \nl
  17 & 20.69 & 0.17 &  19.73 & 0.12 &  18.97 & 0.28 & 0.96 & 0.76 & -2.1 &  1.2  \nl
  18 & 20.71 & 0.18 &  19.83 & 0.13 & $>$19.06 & 0.00 & 0.88 & 0.00 & -2.1 &  0.8  \nl
  19 & 20.84 & 0.20 & $>$20.85 & 0.00 & $>$19.08 & 0.00 & 0.00 & 0.00 & -1.3 & -1.1   \nl
  20 & 20.92 & 0.21 &  20.36 & 0.21 & $>$19.03 & 0.00 & 0.57 & 0.00 &  1.5 &  0.2  \nl
  21 & 20.99 & 0.22 &  20.29 & 0.20 & $>$19.09 & 0.00 & 0.70 & 0.00 &  0.5 &  1.7  \nl
 22 & $>$21.38 & 0.00 &  20.49 & 0.23 & $>$19.09 & 0.00 & 0.00 & 0.00 & -2.0 &  1.5  \nl

\hline

\multicolumn{11}{c}{NGC 2623}\nl
\hline

  01 & 19.50 & 0.07 &  18.84 & 0.06 &  18.84 & 0.25 & 0.66 & 0.00 & -7.9 &  1.2  \nl
  02 & 20.27 & 0.15 &  19.66 & 0.13 & $>$19.10 & 0.00 & 0.61 & 0.00 & -1.2 & -2.4  \nl
  03 & 20.84 & 0.24 &  19.47 & 0.11 & $>$19.10 & 0.00 & 1.37 & 0.00 &  1.4 &  1.4  \nl
  04 & 20.99 & 0.27 &  19.86 & 0.15 &  18.35 & 0.17 & 1.14 & 1.51 &  0.3 &  2.6  \nl
 05 & $>$21.15 & 0.00 &  20.28 & 0.21 &  18.37 & 0.17 & 0.00 & 1.91 & -1.2 &  1.7  \nl
\hline

\multicolumn{11}{c}{IC 883}\nl
\hline

  01 & 18.02 & 0.02 &  17.56 & 0.02 &  17.62 & 0.10 & 0.45 & -.06 &  4.8 &  7.7  \nl 
  02 & 18.31 & 0.03 &  16.34 & 0.01 &  15.41 & 0.01 & 1.97 & 0.93 & -0.4 & -0.7  \nl
  03 & 18.34 & 0.03 &  17.27 & 0.02 &  16.88 & 0.05 & 1.07 & 0.39 &  1.2 &  1.7  \nl
  04 & 19.07 & 0.06 &  18.75 & 0.07 &  18.57 & 0.22 & 0.32 & 0.18 & -2.8 & -0.2  \nl
  05 & 19.09 & 0.06 &  18.50 & 0.05 &  18.17 & 0.16 & 0.59 & 0.33 &  2.0 & -0.1  \nl
  06 & 20.56 & 0.21 &  19.53 & 0.13 &  18.75 & 0.26 & 1.04 & 0.78 & -0.5 & -1.5  \nl
  07 & 20.75 & 0.24 &  19.98 & 0.19 & $>$18.98 & 0.00 & 0.77 & 0.00 &  1.3 & -2.2  \nl
  08 & 20.76 & 0.24 &  20.25 & 0.24 & $>$18.95 & 0.00 & 0.50 & 0.00 &  7.2 & -1.5  \nl
  09 & 20.93 & 0.28 &  20.19 & 0.23 & $>$18.95 & 0.00 & 0.74 & 0.00 &  1.4 & -2.9  \nl
 10 & $>$21.06 & 0.00 &  19.16 & 0.09 &  18.09 & 0.15 & 0.00 & 1.07 &  1.7 &  2.3  \nl
 11 & $>$21.07 & 0.00 &  20.40 & 0.27 & $>$18.95 & 0.00 & 0.00 & 0.00 &  0.8 &  1.6  \nl
 12 & $>$21.07 & 0.00 &  20.44 & 0.28 & $>$18.92 & 0.00 & 0.00 & 0.00 & -2.9 & -2.1  \nl

\hline
\nl
\multicolumn{11}{c}{NGC 7469}\nl
\hline

  01 & 16.76 & 0.01 &  16.34 & 0.01 &  15.82 & 0.02 & 0.42 & 0.53 & -1.6 &  0.6  \nl
  02 & 17.41 & 0.01 &  16.93 & 0.01 &  16.74 & 0.04 & 0.48 & 0.20 &  1.0 & -1.3  \nl
  03 & 17.58 & 0.01 &  17.21 & 0.02 &  16.19 & 0.03 & 0.37 & 1.02 &  1.2 & -1.1  \nl
  04 & 17.91 & 0.02 &  18.17 & 0.04 &  17.54 & 0.09 & -.26 & 0.62 &  1.7 &  0.0  \nl
  05 & 18.79 & 0.03 &  18.64 & 0.05 &  19.03 & 0.29 & 0.15 & -.38 &  9.3 & -3.5  \nl
  06 & 19.10 & 0.05 & $>$20.30 & 0.00 &  18.60 & 0.23 & 0.00 & 0.00 & -0.5 &  1.6  \nl
  07 & 19.98 & 0.11 &  19.41 & 0.12 &  18.57 & 0.22 & 0.58 & 0.83 & -2.6 & -2.6  \nl
  08 & 20.01 & 0.11 &  19.27 & 0.10 & $>$18.98 & 0.00 & 0.74 & 0.00 & -3.2 & -1.7  \nl
  09 & 20.87 & 0.24 &  20.06 & 0.21 & $>$18.97 & 0.00 & 0.81 & 0.00 & -9.2 & -0.9  \nl
  10 & 21.10 & 0.29 &  19.99 & 0.19 & $>$18.96 & 0.00 & 1.11 & 0.00 & -3.0 & -2.0  \nl

\hline

\multicolumn{11}{c}{VV 114E}\nl
\hline

  01 & 18.30 & 0.04 &  18.09 & 0.06 &  18.14 & 0.22 & 0.21 & -.05 & -3.2 &  1.5  \nl
  02 & 18.79 & 0.06 &  17.82 & 0.04 &  17.47 & 0.12 & 0.96 & 0.35 & -4.1 &  2.0  \nl
  03 & 18.87 & 0.06 &  16.31 & 0.01 &  15.31 & 0.02 & 2.56 & 1.00 &  1.3 & -0.9  \nl
  04 & 18.88 & 0.06 &  17.17 & 0.02 &  16.21 & 0.04 & 1.71 & 0.96 & -0.5 &  0.7  \nl
  05 & 19.00 & 0.07 &  17.63 & 0.04 &  16.66 & 0.06 & 1.37 & 0.97 &  1.7 & -0.8  \nl
  06 & 19.69 & 0.13 &  18.42 & 0.07 & $>$18.54  & 0.00 & 1.27 & 0.00 &  1.0 & -0.8  \nl
  07 & 19.69 & 0.13 &  19.44 & 0.18 & $>$18.54 & 0.00 & 0.25 & 0.00 &  2.5 & -1.3  \nl
  08 & 19.72 & 0.13 &  18.65 & 0.09 &  18.00 & 0.20 & 1.06 & 0.65 &  2.2 & -1.2  \nl
  09 & 19.78 & 0.14 &  19.37 & 0.17 & $>$18.54 & 0.00 & 0.41 & 0.00 & -0.7 &  1.5  \nl
  10 & 19.83 & 0.14 &  19.17 & 0.14 & $>$18.50 & 0.00 & 0.67 & 0.00 &  2.0 & -1.0  \nl
  11 & 20.03 & 0.17 &  18.89 & 0.11 &  18.20 & 0.23 & 1.13 & 0.69 &  1.8 & -0.1  \nl
  12 & 20.25 & 0.20 &  19.53 & 0.19 &  18.49 & 0.30 & 0.72 & 1.04 &  1.0 &  0.3  \nl
  13 & 20.29 & 0.21 &  19.52 & 0.19 & $>$18.54 & 0.00 & 0.77 & 0.00 &  0.4 &  1.3  \nl
  14 & 20.45 & 0.24 &  19.02 & 0.12 & $>$18.55 & 0.00 & 1.43 & 0.00 &  2.7 & -1.7   \nl
  15 & 20.67 & 0.29 &  19.89 & 0.26 & $>$18.52 & 0.00 & 0.78 & 0.00 &  2.1 & -2.4  \nl
  16 & 20.75 & 0.31 & $>$20.09 & 0.00 & $>$18.36 & 0.00 & 0.00 & 0.00 & -1.3 &  1.5  \nl
 17 & $>$20.75 & 0.00 &  19.73 & 0.23 & $>$18.54 & 0.00 & 0.00 & 0.00 &  2.1 & -1.8  \nl
 18 & $>$20.76 & 0.00 &  19.81 & 0.25 & $>$18.53 & 0.00 & 0.00 & 0.00 &  3.0 & -2.6  \nl
 19 & $>$20.76 & 0.00 &  20.03 & 0.29 & $>$18.50 & 0.00 & 0.00 & 0.00 &  0.5 &  1.1  \nl
\hline

\multicolumn{11}{c}{VV 114W}\nl
\hline

  01 & 18.15 & 0.03 &  17.60 & 0.03 &  17.57 & 0.14 & 0.55 & 0.03 &  0.1 &  0.1  \nl
  02 & 18.53 & 0.05 &  17.90 & 0.04 &  17.84 & 0.17 & 0.63 & 0.06 &  2.7 &  5.3  \nl
  03 & 18.78 & 0.06 &  19.11 & 0.13 & $>$18.53 & 0.00 & -.33 & 0.00 & -3.7 &  6.4  \nl
  04 & 18.84 & 0.06 &  18.50 & 0.08 &  18.10 & 0.22 & 0.34 & 0.40 & -0.9 & -0.3  \nl
  05 & 19.06 & 0.07 &  18.28 & 0.06 &  17.90 & 0.19 & 0.78 & 0.37 &  0.8 &  3.6  \nl
  06 & 19.09 & 0.08 &  19.38 & 0.16 &  18.53 & 0.31 & -.29 & 0.85 &  2.2 &  5.7  \nl
  07 & 19.39 & 0.10 &  19.16 & 0.14 & $>$18.51 & 0.00 & 0.23 & 0.00 & -0.2 & -0.5  \nl
  08 & 19.45 & 0.11 &  19.31 & 0.16 & $>$18.52 & 0.00 & 0.15 & 0.00 &  1.9 & -2.3  \nl
  09 & 19.71 & 0.13 &  18.93 & 0.11 & $>$18.53 & 0.00 & 0.78 & 0.00 &  1.6 & -0.6  \nl
  10 & 19.75 & 0.14 &  19.54 & 0.19 & $>$18.51 & 0.00 & 0.21 & 0.00 &  3.3 &  4.8  \nl
  11 & 19.92 & 0.16 &  19.80 & 0.24 & $>$18.48 & 0.00 & 0.12 & 0.00 & -3.7 &  2.8  \nl
  12 & 19.92 & 0.16 &  19.90 & 0.26 & $>$18.50 & 0.00 & 0.03 & 0.00 &  7.0 &  4.6  \nl
  13 & 20.00 & 0.17 &  19.31 & 0.16 & $>$18.51 & 0.00 & 0.69 & 0.00 &  2.3 &  5.4  \nl
  14 & 20.11 & 0.19 &  19.49 & 0.18 & $>$18.51 & 0.00 & 0.63 & 0.00 &  2.5 &  1.0  \nl
  15 & 20.12 & 0.19 & $>$20.14 & 0.00 & $>$18.47 & 0.00 & 0.00 & 0.00 &  0.6 & -0.1  \nl
  16 & 20.16 & 0.20 &  19.76 & 0.23 & $>$18.51 & 0.00 & 0.40 & 0.00 &  4.3 &  3.5  \nl
  17 & 20.24 & 0.21 &  19.28 & 0.15 & $>$18.52 & 0.00 & 0.96 & 0.00 &  2.2 &  2.3  \nl
  18 & 20.50 & 0.26 &  19.85 & 0.25 & $>$18.51 & 0.00 & 0.65 & 0.00 &  4.9 & -0.5  \nl
  19 & 20.52 & 0.26 &  19.80 & 0.24 & $>$18.50 & 0.00 & 0.73 & 0.00 &  1.8 &  4.8  \nl
  20 & 20.57 & 0.28 &  19.72 & 0.22 & $>$18.50 & 0.00 & 0.85 & 0.00 &  3.7 &  2.4  \nl
  21 & 20.58 & 0.28 & $>$20.13 & 0.00 & $>$18.49 & 0.00 & 0.00 & 0.00 & -2.0 &  7.1  \nl
  22 & 20.59 & 0.28 &  19.84 & 0.24 & $>$18.52 & 0.00 & 0.75 & 0.00 &  2.2 &  4.7  \nl
  23 & 20.59 & 0.28 &  20.13 & 0.31 & $>$18.50 & 0.00 & 0.46 & 0.00 &  6.8 &  1.7  \nl
  24 & 20.65 & 0.29 &  19.65 & 0.21 & $>$18.50 & 0.00 & 1.00 & 0.00 &  1.6 &  1.8  \nl
  25 & 20.69 & 0.30 & $>$20.12 & 0.00 & $>$18.50 & 0.00 & 0.00 & 0.00 &  3.9 &  0.9  \nl
 26 & $>$20.53 & 0.00 &  19.93 & 0.26 & $>$18.39 & 0.00 & 0.00 & 0.00 &  2.9 &  4.9  \nl
 27 & $>$20.72 & 0.00 &  19.75 & 0.23 & $>$18.51 & 0.00 & 0.00 & 0.00 &  2.0 &  2.3  \nl
\hline

\multicolumn{11}{c}{NGC 6240S}\nl
\hline

  01 & 18.93 & 0.09 &  18.37 & 0.08 & $>$18.44 & 0.00 & 0.57 & 0.00 & -0.6 & -6.5  \nl
  02 & 19.21 & 0.11 &  18.61 & 0.10 & $>$18.44 & 0.00 & 0.59 & 0.00 & -3.0 &  6.4  \nl
\hline

\multicolumn{11}{c}{VIIZw031}\nl
\hline

  01 & 19.20 & 0.04 &  18.09 & 0.02 &  17.51 & 0.08 & 1.11 & 0.58 & -0.8 &  0.0  \nl
  02 & 19.30 & 0.04 &  18.48 & 0.03 &  17.31 & 0.07 & 0.83 & 1.17 & -0.1 & -0.7  \nl
  03 & 19.41 & 0.05 &  18.57 & 0.03 &  18.19 & 0.14 & 0.84 & 0.38 & -0.9 & -1.3  \nl
  04 & 19.55 & 0.05 &  18.58 & 0.03 &  18.02 & 0.12 & 0.97 & 0.56 &  0.7 &  0.9  \nl
  05 & 19.73 & 0.06 &  18.51 & 0.03 &  18.17 & 0.14 & 1.22 & 0.34 &  1.4 &  0.3  \nl
  06 & 19.86 & 0.07 &  19.46 & 0.07 &  18.41 & 0.17 & 0.40 & 1.05 & -0.9 & -1.0  \nl
  07 & 19.89 & 0.07 &  18.72 & 0.04 &  18.05 & 0.13 & 1.18 & 0.66 &  1.0 & -0.4  \nl
  08 & 19.95 & 0.07 &  18.21 & 0.02 &  17.86 & 0.11 & 1.74 & 0.34 &  0.5 &  0.1  \nl
  09 & 19.98 & 0.07 &  19.50 & 0.07 & $>$19.10 & 0.00 & 0.47 & 0.00 &  0.7 & -0.7  \nl
  10 & 20.06 & 0.08 &  19.30 & 0.06 &  18.90 & 0.26 & 0.77 & 0.40 & -2.5 & -1.8  \nl
  11 & 20.18 & 0.09 &  19.18 & 0.06 &  18.52 & 0.19 & 1.00 & 0.66 & -2.8 & -1.8  \nl
  12 & 20.20 & 0.09 &  18.91 & 0.04 &  18.42 & 0.18 & 1.30 & 0.48 & -1.0 &  0.3  \nl
  13 & 20.29 & 0.10 &  19.26 & 0.06 &  18.60 & 0.21 & 1.02 & 0.66 & -0.8 &  1.3  \nl
  14 & 20.31 & 0.10 &  19.21 & 0.06 &  18.61 & 0.21 & 1.10 & 0.61 & -1.1 & -0.6  \nl 
  15 & 20.41 & 0.11 &  18.76 & 0.04 &  17.78 & 0.10 & 1.65 & 0.97 &  1.3 &  0.0  \nl
  16 & 20.47 & 0.12 &  18.89 & 0.04 &  18.20 & 0.15 & 1.58 & 0.69 & -1.0 & -0.3  \nl
  17 & 20.64 & 0.13 &  19.17 & 0.05 & $>$19.11 & 0.00 & 1.47 & 0.00 & -0.3 &  0.5   \nl
  18 & 20.68 & 0.14 &  19.29 & 0.06 &  18.70 & 0.22 & 1.39 & 0.59 & -1.9 & -0.1  \nl
  19 & 20.71 & 0.14 &  19.91 & 0.10 & $>$19.10 & 0.00 & 0.80 & 0.00 & -0.7 &  0.8  \nl
  20 & 20.88 & 0.17 &  19.68 & 0.09 & $>$19.06 & 0.00 & 1.20 & 0.00 & -0.7 &  0.6  \nl
  21 & 21.39 & 0.26 &  20.11 & 0.13 & $>$19.09 & 0.00 & 1.28 & 0.00 &  0.4 & -1.3  \nl
  22 & 21.45 & 0.27 &  19.71 & 0.09 &  18.45 & 0.18 & 1.73 & 1.27 & -1.1 & -0.2  \nl
  23 & 21.45 & 0.27 &  20.92 & 0.25 & $>$18.95 & 0.00 & 0.53 & 0.00 & -0.5 &  3.2  \nl
  24 & 21.59 & 0.30 &  19.43 & 0.07 &  18.97 & 0.28 & 2.16 & 0.47 & -0.3 &  0.8  \nl
  25 & 21.60 & 0.30 &  20.07 & 0.12 & $>$19.10 & 0.00 & 1.53 & 0.00 & -1.0 & -0.8  \nl
  26 & 21.62 & 0.31 &  20.60 & 0.19 & $>$18.85 & 0.00 & 1.03 & 0.00 &  0.5 & -2.8  \nl
 27 & $>$21.63 & 0.00 &  21.07 & 0.28 & $>$19.05 & 0.00 & 0.00 & 0.00 & -0.6 & -3.1  \nl
 28 & $>$21.63 & 0.07 &  19.97 & 0.11 & $>$19.11 & 0.00 & 0.00 & 0.00 &  0.4 & -0.6  \nl
\hline

\multicolumn{11}{c}{IR 15250+3609}\nl
\hline

  01 & 18.49 & 0.04 &  17.81 & 0.04 &  17.92 & 0.16 & 0.67 & -.11 &  0.5 &  0.4  \nl
  02 & 19.70 & 0.11 &  18.99 & 0.10 & $>$18.71 & 0.00 & 0.71 & 0.00 & -0.3 &  0.4  \nl
  03 & 20.37 & 0.19 &  19.85 & 0.22 & $>$18.68 & 0.00 & 0.53 & 0.00 &  2.0 & -1.0  \nl
  04 & 20.61 & 0.24 & $>$20.29 & 0.00 & $>$18.68 & 0.00 & 0.00 & 0.00 &  0.8 &  0.8  \nl
\hline

\multicolumn{11}{c}{UGC 5101}\nl
\hline

  01 & 19.60 & 0.05 &  18.74 & 0.04 & $>$18.57 & 0.00 & 0.86 & 0.00 & -0.7 & -0.4  \nl
  02 & 20.52 & 0.12 &  19.70 & 0.09 & $>$19.21 & 0.00 & 0.82 & 0.00 &  3.4 & -0.8  \nl
 03 & $>$21.66 & 0.00 &  20.25 & 0.15 & $>$19.20 & 0.00 & 0.00 & 0.00 &  1.6 & -2.0  \nl
\hline

\multicolumn{11}{c}{IR 10565+2448W}\nl
\hline

  01 & 18.42 & 0.02 &  17.98 & 0.02 &  17.41 & 0.06 & 0.45 & 0.56 &  0.4 & -0.3  \nl
  02 & 19.01 & 0.03 &  18.71 & 0.04 &  17.31 & 0.06 & 0.30 & 1.40 & -1.1 &  0.3  \nl
  03 & 21.15 & 0.22 &  20.51 & 0.21 & $>$19.20 & 0.00 & 0.64 & 0.00 & -0.4 & -4.1  \nl
\hline

\multicolumn{11}{c}{IR 22491$-$1808W}\nl
\hline

  01 & 19.36 & 0.05 &  18.44 & 0.03 &  18.19 & 0.13 & 0.92 & 0.26 &  1.7 & -0.5  \nl
  02 & 19.87 & 0.08 &  19.15 & 0.06 &  18.74 & 0.21 & 0.72 & 0.41 & -0.7 & -0.6  \nl
  03 & 20.49 & 0.14 &  19.90 & 0.12 & $>$19.21 & 0.00 & 0.60 & 0.00 &  2.4 &  1.9  \nl
  04 & 21.01 & 0.21 &  19.80 & 0.11 & $>$19.22 & 0.00 & 1.21 & 0.00 & -1.3 & -1.1  \nl
  05 & 21.20 & 0.25 &  20.41 & 0.19 & $>$19.20 & 0.00 & 0.79 & 0.00 &  2.0 &  1.3  \nl
 06 & $>$21.47 & 0.00 &  20.33 & 0.18 & $>$19.20 & 0.00 & 0.00 & 0.00 &  1.4 &  0.1  \nl
\hline

\multicolumn{11}{c}{Mrk 273S}\nl
\hline

  01 & 19.49 & 0.10 &  19.89 & 0.22 &  18.13 & 0.21 & -.40 & 1.76 &  1.0 & -0.2  \nl
\hline

\multicolumn{11}{c}{Arp 220W}\nl
\hline

  01 & 18.99 & 0.03 &  17.61 & 0.02 &  17.16 & 0.05 & 1.38 & 0.45 &  0.2 &  0.7  \nl
  02 & 20.30 & 0.09 &  19.42 & 0.08 &  19.29 & 0.30 & 0.89 & 0.13 & -1.5 &  4.8  \nl
  03 & 21.25 & 0.20 &  20.67 & 0.25 & $>$19.30 & 0.00 & 0.58 & 0.00 &  4.7 &  2.3  \nl
  04 & 21.59 & 0.27 & $>$20.95 & 0.00 & $>$19.30 & 0.00 & 0.00 & 0.00 & -0.7 &  2.3  \nl
  05 & 21.61 & 0.28 & $>$20.95 & 0.00 & $>$19.15 & 0.00 & 0.00 & 0.00 & -5.2 & -4.5  \nl
\hline

\multicolumn{11}{c}{IR 17208-0014}\nl
\hline

  01 & 19.24 & 0.07 &  17.76 & 0.03 &  16.96 & 0.07 & 1.48 & 0.80 &  0.3 &  0.7  \nl
  02 & 19.44 & 0.08 &  18.63 & 0.07 &  17.91 & 0.15 & 0.80 & 0.72 &  0.9 & -0.6  \nl
  03 & 19.48 & 0.09 &  18.43 & 0.06 &  17.90 & 0.15 & 1.05 & 0.53 &  1.2 & -1.0  \nl
  04 & 19.57 & 0.09 &  18.81 & 0.08 & $>$18.74 & 0.00 & 0.76 & 0.00 &  0.2 & -1.2  \nl
  05 & 19.66 & 0.10 &  19.20 & 0.12 & $>$18.76 & 0.00 & 0.46 & 0.00 &  1.0 & -0.2  \nl
  06 & 19.94 & 0.13 &  19.28 & 0.12 & $>$18.76 & 0.00 & 0.66 & 0.00 &  0.7 & -0.7  \nl
  07 & 20.01 & 0.14 &  19.24 & 0.12 & $>$18.77 & 0.00 & 0.78 & 0.00 &  1.1 & -1.2  \nl
  08 & 20.13 & 0.15 &  19.32 & 0.13 &  18.60 & 0.27 & 0.81 & 0.73 &  0.9 & -1.5  \nl 
  09 & 20.22 & 0.17 &  17.86 & 0.03 &  16.56 & 0.05 & 2.36 & 1.30 &  0.4 &  0.2  \nl
  10 & 20.25 & 0.17 &  19.63 & 0.17 & $>$18.78 & 0.00 & 0.62 & 0.00 & -0.7 & -0.8  \nl
  11 & 20.62 & 0.24 &  20.04 & 0.24 & $>$18.76 & 0.00 & 0.59 & 0.00 & -0.1 & -0.7   \nl
  12 & 20.65 & 0.24 & $>$20.29 & 0.00 & $>$18.77 & 0.00 & 0.00 & 0.00 & -0.7 & -0.6  \nl
  13 & 20.82 & 0.28 &  20.08 & 0.25 & $>$18.74 & 0.00 & 0.73 & 0.00 &  2.4 & -2.5  \nl
  14 & 20.86 & 0.29 &  20.26 & 0.28 & $>$18.75 & 0.00 & 0.61 & 0.00 & -1.3 & -0.6  \nl
 15 & $>$20.95 & 0.00 &  19.06 & 0.10 &  18.29 & 0.21 & 0.00 & 0.77 & -0.6 & -0.3  \nl
 16 & $>$20.95 & 0.00 &  19.36 & 0.13 & $>$18.75 & 0.00 & 0.00 & 0.00 &  0.8 &  0.7  \nl
 17 & $>$20.96 & 0.00 &  19.84 & 0.20 & $>$18.78 & 0.00 & 0.00 & 0.00 &  8.5 & -5.0  \nl
 18 & $>$20.96 & 0.00 &  20.11 & 0.25 & $>$18.76 & 0.00 & 0.00 & 0.00 & -1.3 & -0.3  \nl
\enddata
\tablenotetext{a}{Upper limits on all magnitudes are 3$\sigma$ rms upper limits.}
\tablenotetext{b}{1 Sigma root-mean-square for the corresponding measured magnitude.}
\tablenotetext{c}{Offset in arcseconds from the brightest 2.2$\mu$m peak.}
\end{deluxetable}

\begin{deluxetable}{lcccccc}
\tablenum{8}
\tablewidth{0pt}
\tablecaption{Radial Profile Fits}
\tablehead{
\multicolumn{1}{c}{Galaxy} &
\multicolumn{1}{c}{R$_{inner}$\tablenotemark{a} } &
\multicolumn{1}{c}{R$_{outer}$\tablenotemark{a} } &
\multicolumn{1}{c}{Best Model\tablenotemark{b}} &
\multicolumn{1}{c}{Scale\tablenotemark{c}} &
\multicolumn{1}{c}{$\chi^{2}$}\tablenotemark{d} &
\multicolumn{1}{c}{Ratio}\tablenotemark{e}  \nl
\colhead{} &
\colhead{(pc)} &
\colhead{(pc)} &
\multicolumn{1}{c}{($r^{1/4}$ or exp)} &
\multicolumn{1}{c}{(kpc)} &
\multicolumn{1}{c}{} &
\multicolumn{1}{c}{$\chi^{2}$s}
}
\startdata 
NGC 4418 &  30 & 1100 & ${\bf r^{1/4} }$  &  0.710 &   0.84 &  26 \nl

      Zw049.057 &  60 & 2000 & ${\bf r^{1/4} }$ &  2.180 &   0.38 &   7 \nl

        NGC 6090 & 120 & 3000 & exp  &  0.690 &   2.19 &   2 \nl

        NGC 2623 &  90 & 2900 & ${\bf r^{1/4} }$ &  1.480 &   1.38 &   6 \nl

          IC 883 & 100 & 3800 & ${\bf r^{1/4} }$ &  2.180 &   0.26 &  11 \nl

        NGC 7469 &  70 & 3100 & neither  &     &  &    \nl

         VV114E &  90 & 2300 & {\bf exp } &  0.520 &   0.94 &   3 \nl

         VV114W &  80 & 1700 & exp   &  0.560 &   0.87 &   1 \nl

        NGC 6240 &  95 & 4500 & ${\bf r^{1/4} }$ &  1.340 &   0.27 &  19 \nl

       VIIZw031 & 230 & 7000 & neither  &     & &   \nl

   IRAS 15250+3609 & 250 & 5150 & $r^{1/4}$ &  1.730 &   2.26 &   2 \nl

        UGC 5101 & 180 & 6180 & ${\bf r^{1/4} }$ &  1.420 &   0.34 &  19 \nl

   IRAS 10565+2448 & 250 & 6300 & ${\bf r^{1/4} }$ &  1.150 &   1.03 &  23 \nl

   IRAS 08572+3915 & 260 & 2200 & neither  &    &   &      \nl

   IRAS 05189--2524 & 170 & 5800 & neither &    &   &     \nl

   IRAS 22491--1808 & 300 & 7000 & $r^{1/4}$ &  8.290 &   1.67 &   2 \nl

         Mrk 273 & 170 & 5100 & exp  &  1.060 &   1.25 &   1 \nl

         Arp 220 &  90 & 3400 & ${\bf r^{1/4} }$  &  2.800 &   0.50 &   7 \nl

     PKS 1345+12 & 480 & 6900 & neither  &      &   &    \nl

 IRAS 12112+0305 & 310 & 2000 & $r^{1/4}$ &  1.190 &   4.53 &   3\nl

 IRAS 14348--1447 & 360 & 3000 & ${\bf r^{1/4} }$ &  4.190 &   0.84 &   8 \nl

 IRAS 17208--0014 & 180 & 6600 & $r^{1/4}$ &  2.610 &   2.72 &   2 \nl

   IRAS 07598+6508 & 680 & 6900 & neither &  &    &   \nl

        Mrk 1014 & 700 & 6900 & exp  &  1.980 &   6.22 &   2 \nl
\enddata
\tablenotetext{a}{In all cases, the inner radius for the fit, R$_{inner}$,
is at 0.22\arcsec\ radius to be outside the difraction radius. Usually,
R$_{outer}$, the outer radius for the fit is at the detection edge of
the galaxy, but in a few instances, R$_{outer}$ was reduced to avoid a
secondary nucleus.}

\tablenotetext{b}{Functional form of best fitting model to the 1.6 $\mu$m
image. Entries with bold type are those with excellent fits and where
the fit clearly discriminates the $r^{1/4}$-law and exponential forms.}

\tablenotetext{c}{Scale length for best-fit function. For $r^{1/4}$ ,
this is the radius in which half the total flux is contained; for the
exponential disk, it is the e-folding length in radius.}

\tablenotetext{d}{$\chi^{2}$ normalized by the number of radial bins in
the fit.}

\tablenotetext{e}{The ratio of $\chi^{2}$'s for the worse fitting model
to the better fitting model. This serves as an estimator of how well
the data favors the best-fit model over the other model, i.e. a ratio
of 1 means both fits are equally acceptable and a high ratio implies
that the best-fit funtion is highly favored.}

\end{deluxetable}

\begin{deluxetable}{lcccc}
\tablenum{9}       
\tablewidth{0pt}
\tablecaption{ Half-Light Radii For Flux within 3 kpc}
\tablehead{
\colhead {Name } &
\colhead {IR Class} &
\colhead { $R_{1/2} (1.1\mu$m) } &
\colhead {$R_{1/2} (1.6\mu$m) } &
\colhead {$R_{1/2} (2.2\mu$m) } \nl
\colhead {} &
\colhead {} &
\colhead {kpc} &
\colhead {kpc} &
\colhead {kpc} \nl
}
\startdata
NGC 4418&W& 0.34& 0.32& 0.26\nl
Zw049.057&C& 0.68& 0.62& 0.44\nl
NGC 6090&W& 1.06& 1.12& 1.07\nl
NGC 2623&C& 0.92& 0.73& 0.39\nl
IC 883&C& 1.05& 0.93& 0.62\nl
NGC 7469&W& 0.21& 0.27& 0.27\nl
VV 114E&C& 0.87& 0.79& 0.61\nl
VV 114W&C& 0.86& 0.87& 0.78\nl
NGC 6240&C& 0.79& 0.73& 0.61\nl
VIIZw031&C& 1.35& 1.28& 1.17\nl
IRAS 15250+36&C& 1.09& 1.01& 0.65\nl
UGC 5101&C& 1.01& 0.80& 0.45\nl
IRAS 10565+24&C& 0.89& 0.79& 0.65\nl
IRAS 08572+39&W& 0.83& 0.69& 0.14\nl
IRAS 05189-25&W& 0.14& 0.10& 0.12\nl
IRAS 22491-18&C& 1.60& 1.66& 1.62\nl
Mrk 273&C& 1.31& 1.17& 0.90\nl
Arp 220&C& 1.33& 1.09& 0.58\nl
PKS 1345+12&W& 1.03& 1.08& 0.39\nl
IRAS 12112+0305&C& 0.97& 0.79& 0.45\nl
IRAS 14348-1447&C& 1.23& 1.07& 0.67\nl
IRAS 17208-0014&C& 1.28& 1.10& 0.88\nl
IRAS 07598+65&W& 0.17& 0.22& 0.30\nl
Mrk 1014&W& 0.24& 0.28& 0.36\nl
\enddata
\end{deluxetable}

\end{document}